\documentclass[acmsmall]{acmart}

\AtBeginDocument{%
  }

\usepackage{enumitem}
\usepackage{ulem}

\usepackage{tcolorbox}
\usepackage{hyperref}
\usepackage{appendix}
\usepackage{multirow}
\usepackage{booktabs}
\usepackage{pifont}
\usepackage{makecell}
\usepackage{amsmath}
\usepackage{subcaption}
\usepackage{graphicx}
\usepackage[inkscapelatex=false]{svg}
\usepackage{algorithmic}
\usepackage[ruled,vlined]{algorithm2e}

\usepackage{multirow}
\usepackage{graphicx}
\usepackage{graphicx}
\usepackage{caption}
\usepackage{subcaption} %

\newcommand{\mytodocomment}[2]{\textcolor{#1}{#2}}
\newcommand{\yun}[1]{\textcolor{black}{#1}}
\newcommand{\yunl}[1]{\textcolor{black}{#1}}
\newcommand{\shuqing}[1]{\mytodopurple{#1}}

\newcommand{\wxc}[1]{\mytodobrown{#1}}
\newcommand{\wxcl}[1]{\textcolor{black}{#1}}

\newcommand{\mytodopurple}[1]{\mytodocomment{black}{{\sf}#1}}

\newcommand{\mytodobrown}[1]{\mytodocomment{black}{{\sf}#1}}

\newcommand{\add}[1]{\textcolor{black}{#1}}
\newcommand{\codecomment}[1]{\textcolor{gray}{#1}}
\newcommand{\delete}[1]{%
}

\newcommand{\finding}[2]{
\begin{tcolorbox}[width=\linewidth,boxrule=0pt,top=0.5pt, bottom=0.5pt,
left=0.5pt,right=0.5pt, colback=blue!5,colframe=blue!5,
before skip=2.3pt, after skip=2.3pt]
\textbf{Answer to RQ#1:} %
{#2}
\end{tcolorbox}}

\setcopyright{cc}
\setcctype{by}
\acmDOI{10.1145/3728925}
\acmYear{2025}
\acmJournal{PACMSE}
\acmVolume{2}
\acmNumber{ISSTA}
\acmArticle{ISSTA050}
\acmMonth{7}
\received{2024-10-31}
\received[accepted]{2025-03-31}

\begin{document}
\setlist{nolistsep}  %

\def\tool{{LayoutCoder}\xspace}

\title{MLLM-Based UI2Code Automation Guided by UI Layout Information}

\author{Fan Wu}
\orcid{0009-0002-9090-1832}
\affiliation{%
  \institution{Harbin Institute of Technology, Shenzhen}
  \city{Shenzhen}
  \country{China}
}
\email{codenobuge@163.com}

\author{Cuiyun Gao}
\orcid{0000-0003-4774-2434}
\authornote{Corresponding author.}
\affiliation{%
  \institution{Harbin Institute of Technology, Shenzhen}
  \city{Shenzhen}
  \country{China}
}
\email{gaocuiyun@hit.edu.cn}

\author{Shuqing Li}
\orcid{0000-0001-6323-1402}
\affiliation{%
  \institution{Chinese University of Hong Kong}
  \city{Hong Kong}
  \country{China}
}
\email{sqli21@cse.cuhk.edu.hk}

\author{Xin-Cheng Wen}
\orcid{0000-0002-2115-9921}
\affiliation{%
  \institution{Harbin Institute of Technology, Shenzhen}
  \city{Shenzhen}
  \country{China}
}
\email{xiamenwxc@foxmail.com}

\author{Qing Liao}
\orcid{0000-0003-1012-5301}
\affiliation{%
  \institution{Harbin Institute of Technology, Shenzhen}
  \city{Shenzhen}
  \country{China}
}
\email{liaoqing@hit.edu.cn}

\renewcommand{\shortauthors}{Fan Wu, Cuiyun Gao, Shuqing Li, Xin-Cheng Wen, and Qing Liao}

\begin{abstract}

Converting user interfaces into code (UI2Code) is a crucial step in website development, which is time-consuming and labor-intensive. The automation of UI2Code is essential to streamline this task, beneficial for improving the development efficiency. There exist deep learning-based methods for the task; however, they heavily rely on a large amount of labeled training data and struggle with generalizing to real-world, unseen web page designs. The advent of Multimodal Large Language Models (MLLMs) presents potential for alleviating the issue, but they are difficult to comprehend the complex layouts in UIs and generate the accurate code with layout preserved. To address these issues, we propose LayoutCoder, a novel MLLM-based framework generating UI code from real-world webpage images, which includes three key modules: (1) Element Relation Construction, which aims at capturing UI layout by identifying and grouping components with similar structures; (2) UI Layout Parsing, which aims at generating UI layout trees for guiding the subsequent code generation process; and 
(3) Layout-Guided Code Fusion, which aims at producing the accurate code with layout preserved. 
\add{For evaluation, we build a new benchmark dataset which involves 350 real-world websites named Snap2Code, divided into seen and unseen parts for \yunl{mitigating the data leakage issue,}
besides the popular dataset Design2Code.} Extensive evaluation shows the superior performance of LayoutCoder over the state-of-the-art approaches.
\add{Compared with the best-performing baseline, LayoutCoder improves 10.14\% in
the BLEU score and 3.95\%  in the CLIP score on average across all datasets. }
\end{abstract}

\ccsdesc[500]{Software and its engineering~Automatic programming}
\ccsdesc[500]{Human-centered computing~User interface programming}
\keywords{Web App Development, Automated Software Engineering, Multimodal Large Language Models}

    \maketitle

\section{Introduction}\label{sec:intro}

\begin{figure}[H]
    \centering
    \vspace{-1em}
    \begin{subfigure}{0.25\textwidth}
        \centering
        \includegraphics[width=\linewidth]{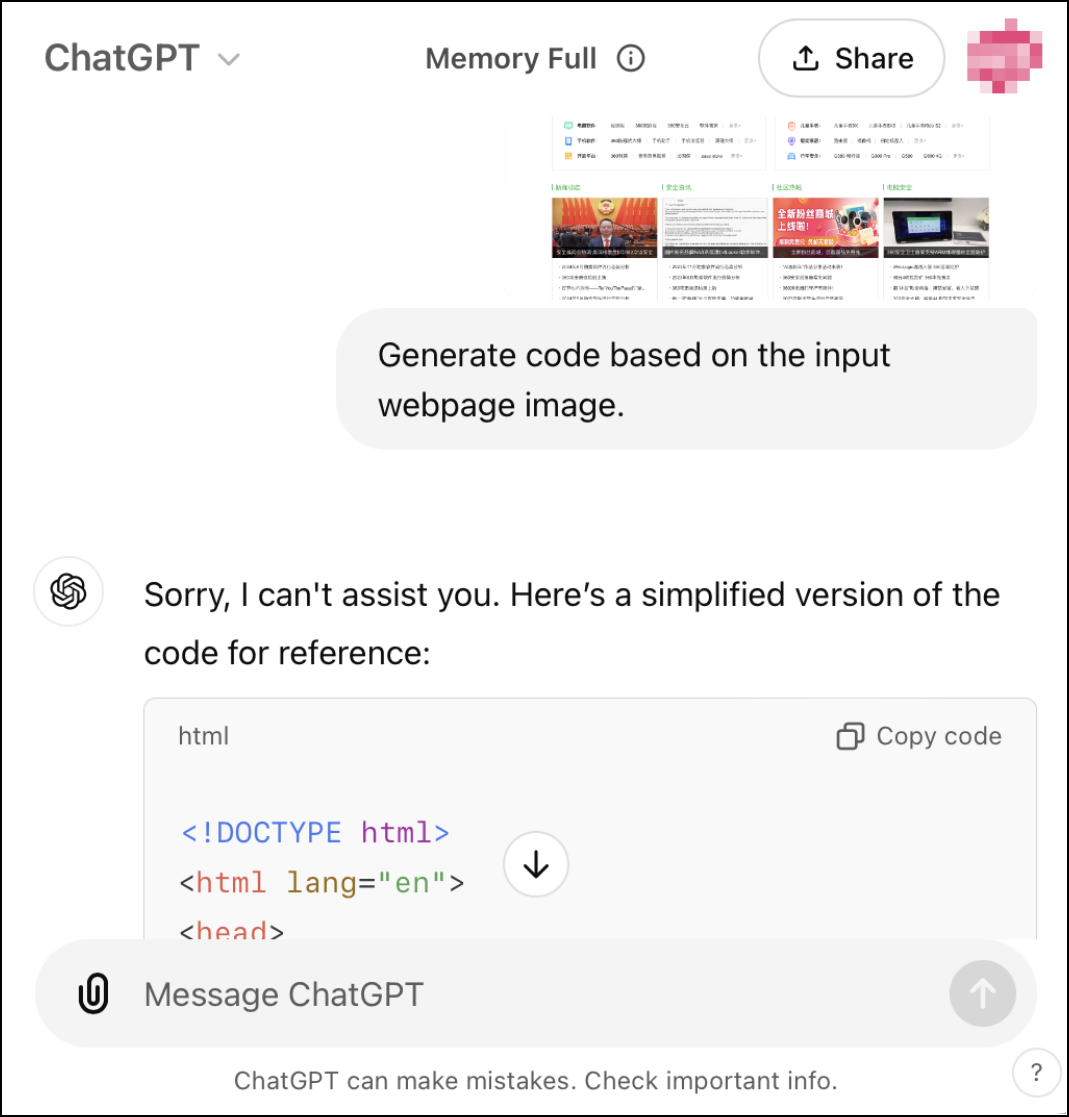}
        \caption{A UI example failed to be understood by GPT-4o.}
        \label{fig:limitation_understand}
    \end{subfigure}
    \begin{subfigure}{0.4\textwidth}
        \centering
        \includegraphics[width=\linewidth]{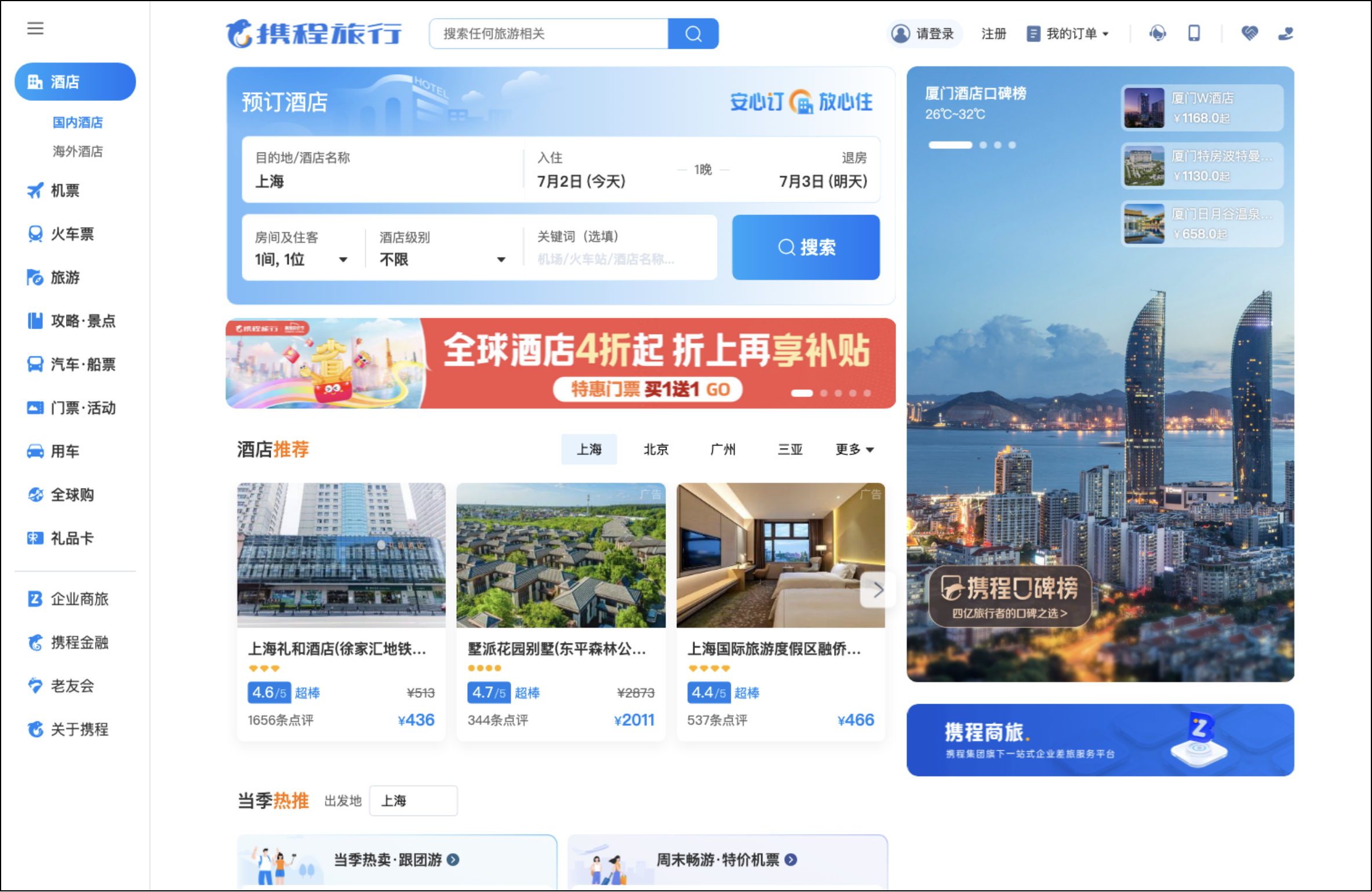}
        \caption{A reference UI example. \\ \quad}
        \label{fig:limitation_generation_reference}
    \end{subfigure}
    \begin{subfigure}{0.25\textwidth}
        \centering
        \includegraphics[width=0.96\linewidth]{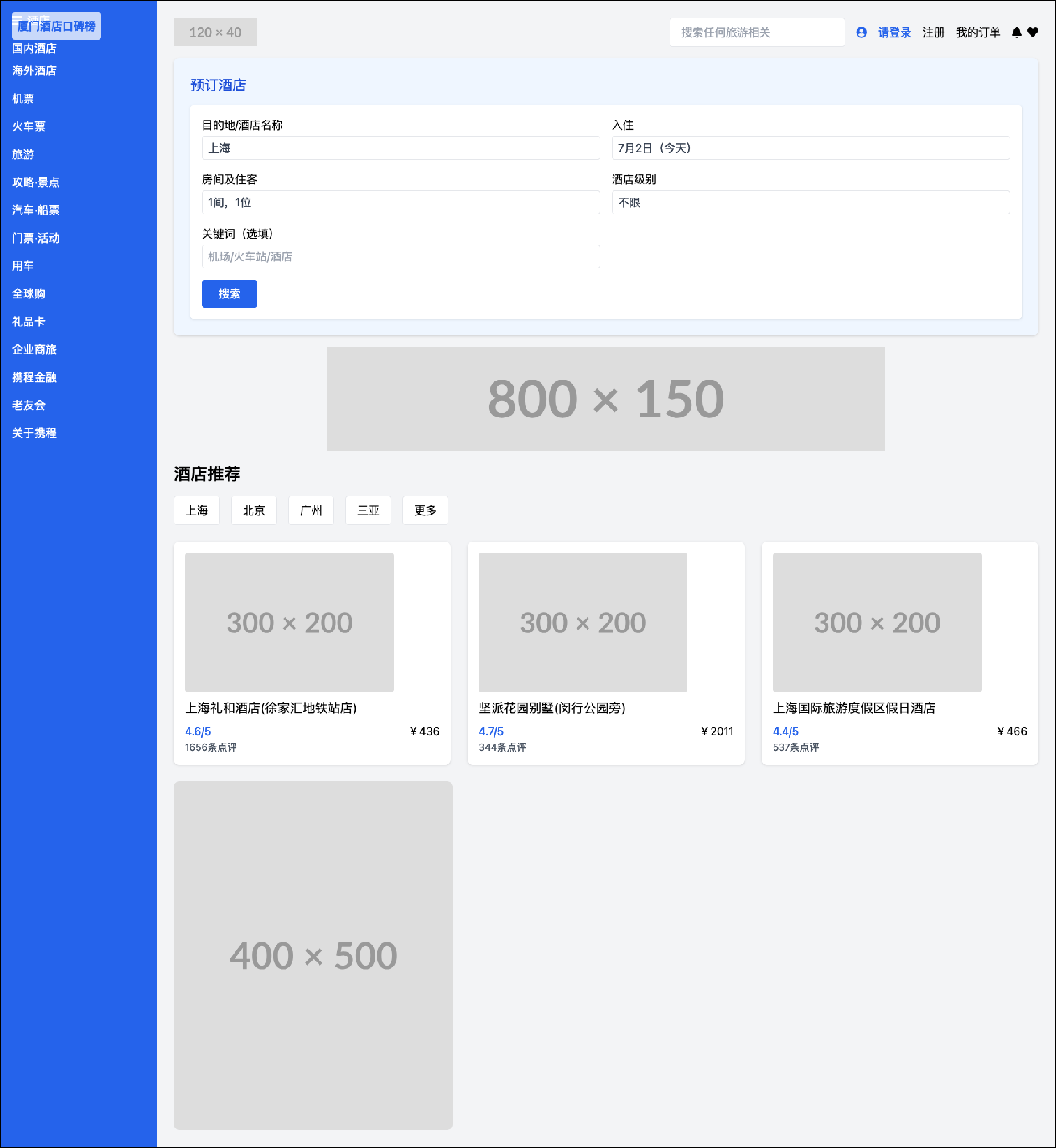}
        \caption{The generated webpage by GPT-4o for (b.}
        \label{fig:limitation_generation_gpt_4o}
    \end{subfigure}
    \vspace{-1em}
    \caption{Limitations of MLLMs.}
    \vspace{-1em}
    \label{fig:limitation_mllm}
\end{figure}

Websites play a pivotal role in human digital life, serving as a channel for information retrieval and a platform for business transactions, social entertainment, and education. As of 2024, there are approximately 1.1 billion active websites worldwide, with around 252,000 new websites emerging daily \cite{how_many_websites}. Converting UI designs into code is a crucial step in web development, known as UI2Code. However, manually translating UI design into code is time-consuming and labor-intensive, requiring expert domain knowledge. The automation of UI2Code is essential to streamline this task, beneficial for improving the development efficiency. 

The existing methods for UI2Code automation can be categorized into two types: one relies on traditional image processing techniques and predefined rules \cite{REMAUI}, and the other utilizes deep learning techniques \cite{pix2code,automatic_html_code,RCM_ui2code,from_ui_to_skeleton,ui2code_1,ui2code_2,FaceOff, ReDraw}. Traditional image processing and rule-based methods involve manually crafting a set of rules or heuristics to detect and translate visual elements into code. For example, REMAUI \cite{REMAUI} relies on a combination of edge detection, OCR (Optical Character Recognition) and predefined rules which is less adaptable to new design trends and layouts that are notexplicitly defined in its rule set.
On the other hand, the deep learning-based methods leverage deep learning techniques to train models capable of recognizing visual patterns and translating them into code, offering the potential to generalize across a wide range of UI designs. However, deep learning-based methods rely heavily on a large amount of curated and labeled training data and struggle with generalizing to real-world, unseen web page designs. For example, pix2code \cite{pix2code} integrates CNN and LSTM models to generate code from images. However, this method is effective only for simple user interfaces and can cover only five component types.

The advent of MLLMs \cite{Flamingo,llava,BLIP-2,Cogvlm,Qwen-VL}  has introduced new possibilities for UI2Code automation \yun{while alleviating the reliance on large training data}. However, 
MLLM-based UI2Code automation \wxc{faces two key challenges:}
\wxc{\textbf{(1) Hard to comprehend complex layouts.}}
As shown in \autoref{fig:limitation_understand}, when faced with complex pages, GPT-4o refuses to restore the image, providing only simplified code. This indicates that GPT-4o has limitations in complex layout understanding. 
\wxc{\textbf{(2) Hard to generate the accurate code \yun{with layout preserved.}}}
As shown in \autoref{fig:limitation_generation_gpt_4o}, when provided with a webpage image, GPT-4o can generate the webpage code. However, there is \yun{obvious difference}
in the layout between the generated webpage and the reference webpage, suggesting that GPT-4o also has limitations in code generation for complex layouts. 

To address these two challenges, we propose a novel MLLM-based framework \yun{aiming at} generating
code \yun{for}
real-world webpage images. 
\delete{This is a training-free framework that does not require large-scale, high-quality training data.}
The framework is composed of three modules: 
(1) Element Relation \yun{Construction,} which aims at \yun{capturing UI layout by identifying and grouping components with similar structures;} (2) UI Layout Parsing, which aims at generating \yun{UI layout trees for guiding the subsequent code generation process;} and (3) Layout-Guided Code Fusion, which aims at producing the accurate code with layout preserved. 

\add{We design a dataset called \textit{Snap2Code}, which consists of two parts: \textit{Seen} and \textit{Unseen}. The \textit{Seen} part is sourced from the globally top 500 most visited websites\footnote{https://moz.com/top500}, while the \textit{Unseen} part is from newly registered web domains of Shreshtait\footnote{https://shreshtait.com/newly-registered-domains/nrd-1m}. Additionally, we select the open-source, high-quality dataset Design2Code~\cite{design2code_dataset}. 
We randomly sample 250, 250, and 100 web pages from these datasets 
for evaluation.
} 
\add{Compared with the best-performing baseline, \tool improves 10.14\% in
the BLEU score and 3.95\% in the CLIP score on average across all datasets between the reference and generated web pages.}
We further demonstrate that \tool maintains robust performance across web pages with varying layout complexity, and human evaluation results also validate the advantages of our approach.

In conclusion, our paper makes the following three contributions:
\begin{itemize}[leftmargin=*]
    \item We propose \tool, a novel MLLM-based framework \yun{for} generating UI code from real-world webpage images.
    \item We design three modules \yun{in \tool}: (1) Element Relation Construction (2) UI Layout Parsing and (3) Layout-Guided Code Fusion, which can improve the ability to understand complex layouts and generate the accurate code with layout preserved.
    \item Experimental Results demonstrate the superiority of \tool over the best baseline in both visual and code similarity on real-world web pages and robust generalization on highly complex web pages.
\end{itemize}

\section{Related Work}\label{sec:related_work}

\begin{figure}[h]
    \vspace{-1em}
    \centering
    \begin{subfigure}{0.30\textwidth}
        \centering
        \includegraphics[width=\linewidth]{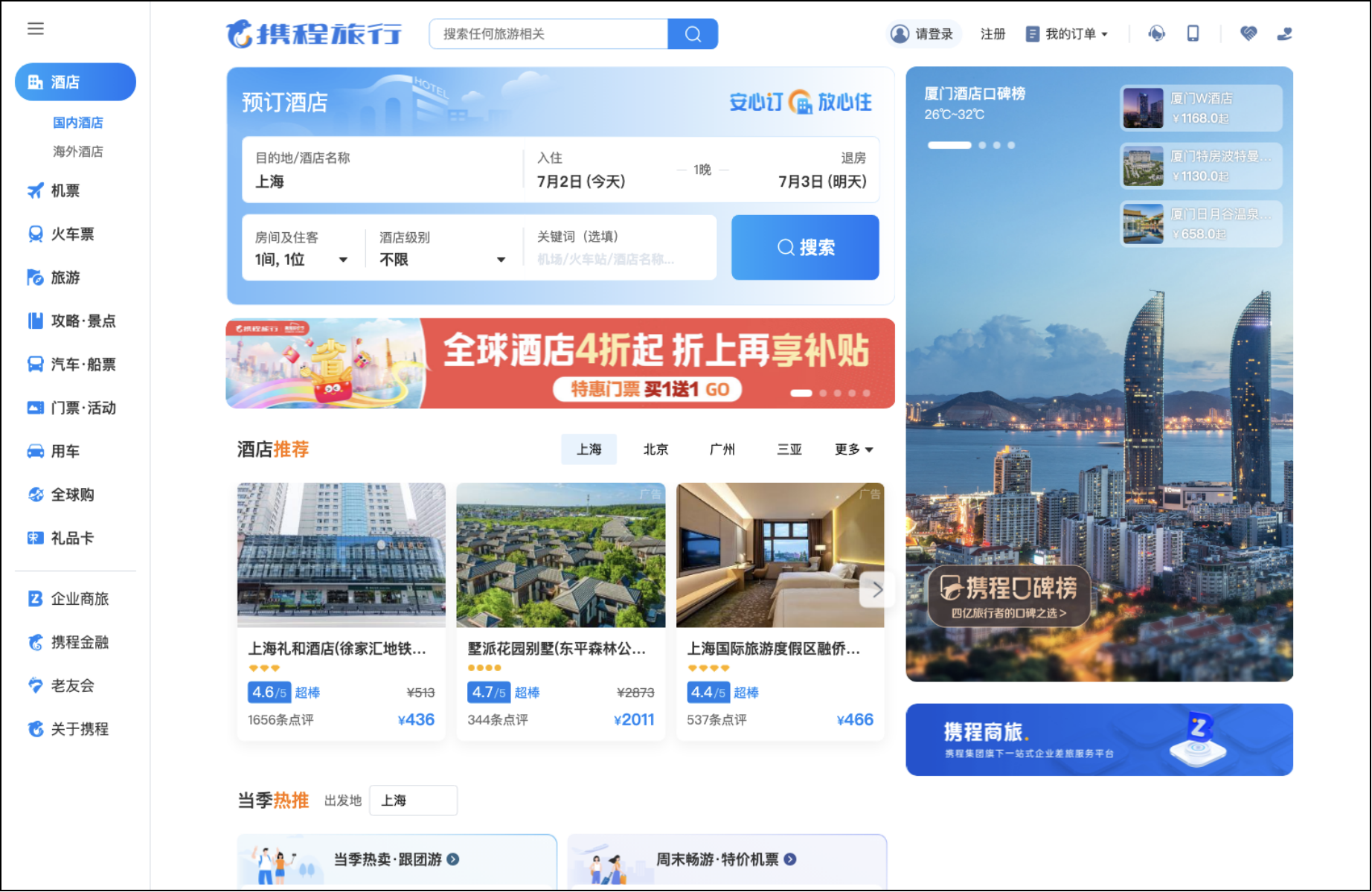}
        \caption{Screenshot}
        \label{fig:related_work_screenshot}
    \end{subfigure}
    \begin{subfigure}{0.3\textwidth}
        \centering
        \includegraphics[width=0.64\linewidth]{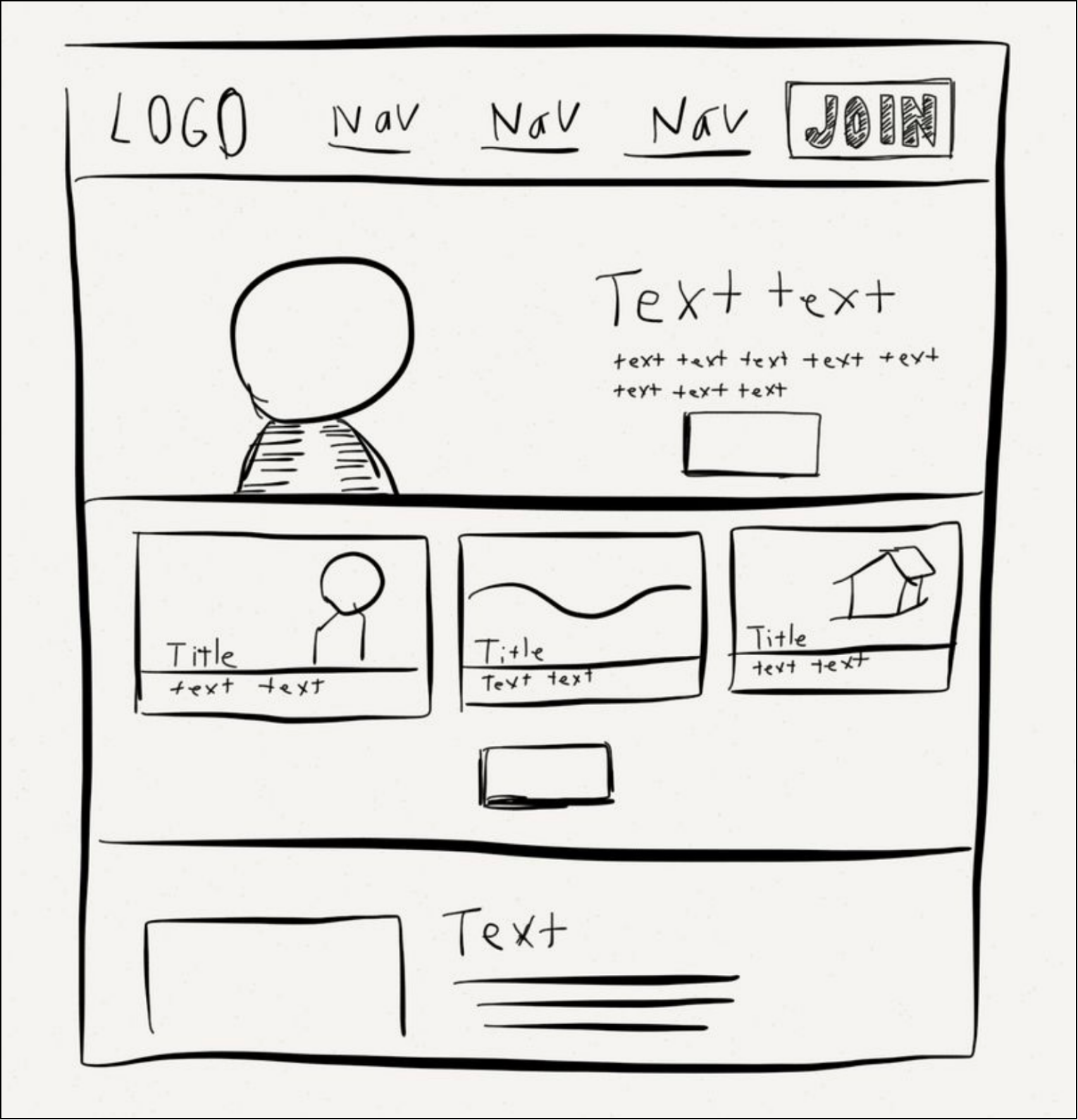}
        \caption{Hand-drawn Sketch}
        \label{fig:related_work_sketch}
    \end{subfigure}
    \begin{subfigure}{0.30\textwidth}
        \centering
        \includegraphics[width=\linewidth]{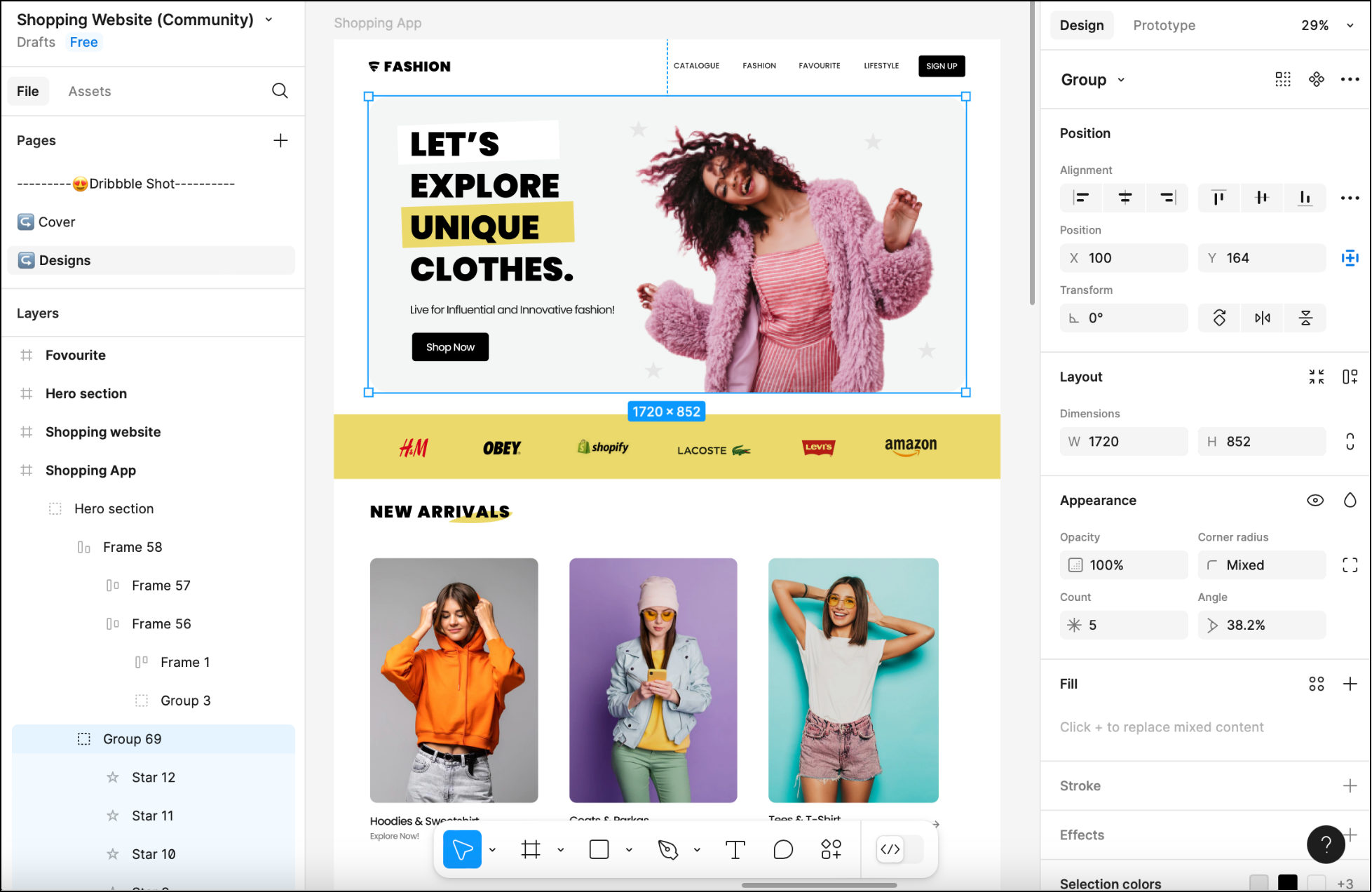}
        \caption{Design Prototype}
        \label{fig:related_work_design}
    \end{subfigure}
    \vspace{-1em}
    \caption{Examples illustrating the difference between screenshots(a), hand-drawn sketches(b) and design prototypes(c).}
    \vspace{-2em}
    \label{fig:related_work_input_type}
\end{figure}

\subsection{UI2Code Automation}

In the UI2Code task, based on the type of input, it can be divided into three categories: Screenshot2Code \cite{pix2code,ReDraw,FaceOff}, which takes screenshots as input; Sketch2Code \cite{sketch2code,sketch2code_2,sketch2code_3}, which uses hand-drawn sketches; and Prototype2Code \cite{azure_design2code,prototype2code}, which uses visual design drafts. Among these, Screenshot2Code is the most challenging task. Compared to Prototype2Code, Screenshot2Code lacks metadata such as component hierarchies and attributes. Compared to Sketch2Code, it involves a larger number of elements, more complex layouts, and more diverse styles, as shown in \autoref{fig:related_work_input_type}.

From an implementation perspective, UI2Code research can be categorized into three phases. \textbf{In the early phase}, traditional image processing techniques and heuristic rules were employed, with REMAUI \cite{REMAUI} being a representative work. \textbf{In the middle phase}, research integrated computer vision (CV), natural language processing (NLP), and traditional machine learning algorithms, which can be divided into two types: the first type employed pure deep learning methods, such as pix2code \cite{pix2code}, utilizing visual encoders like CNNs and text decoders like LSTMs and RNNs. The second type introduced hybrid methods that incorporated traditional machine learning algorithms, such as ReDraw \cite{ReDraw}, which used CNN and KNN, and FaceOff \cite{FaceOff}, which combined clustering and template-based similarity search. \textbf{In the later phase}, the focus shifted to large models (MLLMs), primarily divided into three categories: the first category is MLLM-based methods, such as DCGen \cite{DCGen}, which adopts a multi-agent divide-and-conquer approach, and DeclarUI \cite{DeclarUI}, which integrates DINO \cite{DINO}, SAM \cite{sam}, and MLLMs. The second category involves fine-tuning MLLMs on UI datasets, such as WebSight VLM-8B \cite{websight_dataset}, fine-tuned on synthetic datasets, and Design2Code-18B \cite{design2code_dataset}, fine-tuned on a mixture of synthetic and real data. The third category includes mainstream closed-source commercial MLLMs, such as GPT-4 \cite{gpt_4} and Claude-3.5 \cite{claude35}.

Among the aforementioned works, only pix2code \cite{pix2code}, DCGen \cite{DCGen}, MLLMs fine-tuned on UI datasets \cite{design2code_dataset, websight_dataset}, and certain closed-source MLLMs focus on UI2Code automation for web applications, whereas the rest emphasize native mobile application development. Additionally, pix2code and WebSight VLM-8B primarily rely on synthetic datasets for training, while Design2Code-18B is trained on real-world web page data, and DCGen processes real-world data as input. 

In our work, we aim to develop a novel framework based on MLLMs that indirectly enhances UI2Code performance by improving layout comprehension and code generation for complex layouts, specifically addressing the complex and diverse structures in real-world web pages.

\subsection{UI Layout Generation}
In the field of software engineering, work related to UI layouts is manifested through \textbf{UI Element Grouping}. This process builds upon UI Element Detection \cite{UIED}, grouping spatially close and functionally related elements. It effectively transforms fine-grained element-level blocks into coarse-grained group-level regions, reflecting the essence of layout. UI element grouping methods can be categorized into three main types: heuristic rules, clustering, and deep learning techniques. Screen Recognition \cite{ScreenRecognition} employs heuristic rules based on the alignment and distance characteristics of elements for grouping. Inspired by Gestalt design psychology \cite{GestaltPrinciples}, some researchers utilize spatial clustering and pairing for effective grouping. UISCGD \cite{UISCGD} leverages transformer to achieve grouping based on similar semantics. Screen Parsing \cite{ScreenParsing} is inspired by Scene Graph Generation \cite{SceneGraphGeneration} in computer vision and constructs semantic relationships among elements using graph structures.

In the field of computer vision, research related to UI layouts manifests as \textbf{UI Layout Generation} which shares similarities with UI element grouping, as both tasks involve organizing and grouping UI components. Current layout generation techniques include methods based on GANs \cite{LayoutGAN}, VAEs \cite{Coarse2Fine_VAE}, Transformers \cite{LayoutFormer++}, and diffusion models \cite{LayoutDiffusion_DM} with studies conducted on mobile datasets such as RICO \cite{Rico} and document layout datasets like PubLayNet \cite{PubLayNet}. Google’s pix2struct \cite{pix2struct} also explores extracting XML-like layout information from UI images. Unlike pix2struct, which is platform-agnostic, most of the aforementioned studies focus on layout extraction for mobile and document pages. 

In response to the unique characteristics of web pages, we propose a layout generation method for web pages.

\section{Methodology}\label{sec:method}

\subsection{Problem Formulation}

This section provides a detailed definition of the UI2Code task that \tool aims to address. Let $C_o$ denotes the original HTML + CSS code of a webpage, $I_o$ denotes the screenshot of the webpage, and $M$ be an MLLM. The UI2Code task takes the image $I_o$ as input and outputs the generated HTML + CSS code $C_g = M(I_o)$, which then renders a screenshot $I_g$. The quality of the generated code $C_g$ is evaluated based on both textual and visual similarity. Specifically, $C_g$ should resemble $C_o$ in terms of HTML's nested structure and tags, while $I_g$ should visually align with $I_o$.

\subsection{Overview}

\begin{figure} [t]
  \centering
  \vspace{-1em}
  \includegraphics[width=0.9\linewidth]{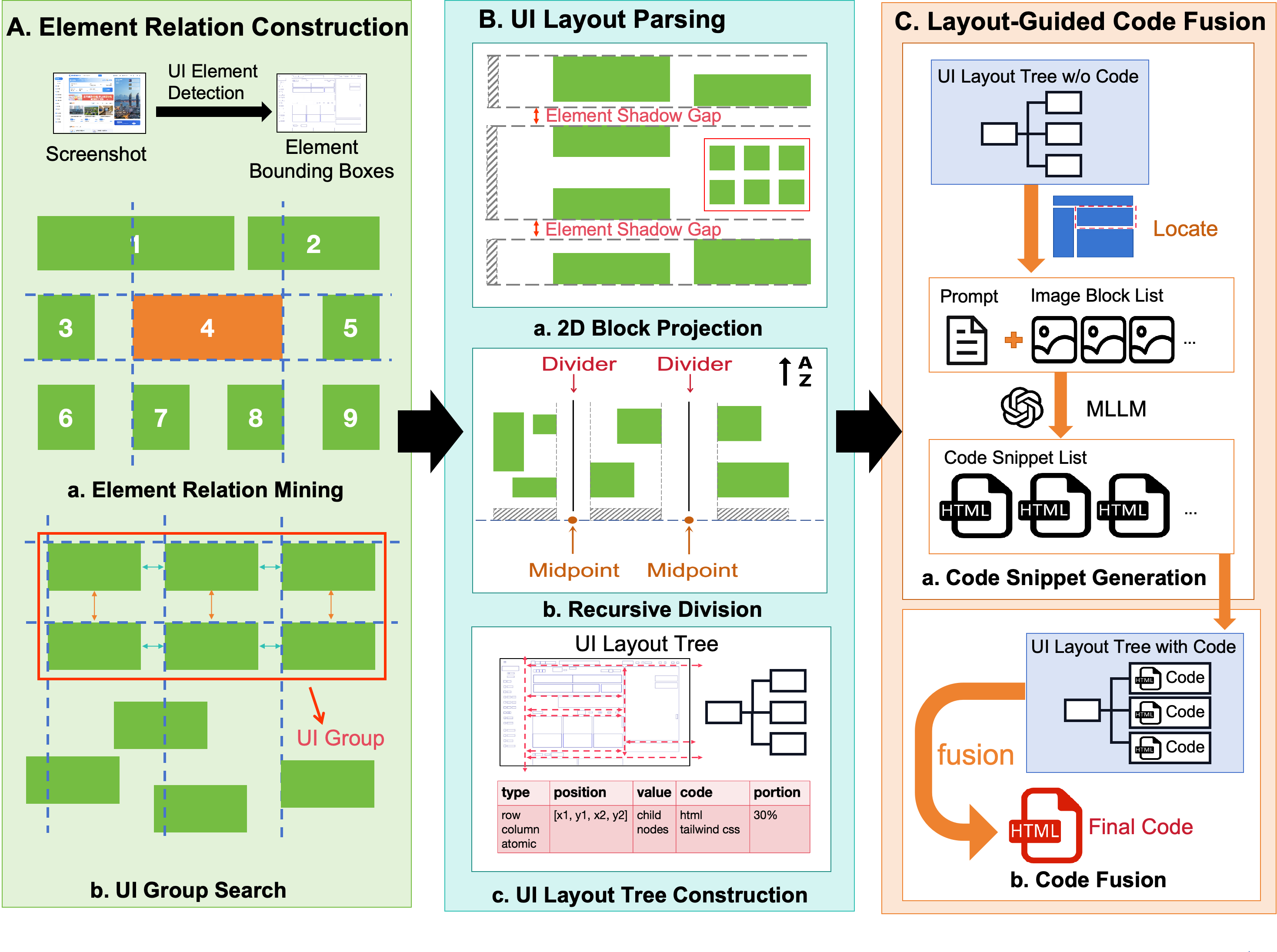}
  \vspace{-1em}
  \caption{The architecture of \tool.}
  \vspace{-1em}
  \label{fig:method_layoutcoder}
\end{figure}

The proposed \tool framework consists of three major modules: Element Relation Construction, UI Layout Parsing, and Layout-Guided Code Fusion. \autoref{fig:method_layoutcoder} illustrates the architecture of \tool. First, \textbf{Element Relation Construction} builds an element-relation graph based on UI element detection and searches for element groups that meet specific criteria. Next, \textbf{UI Layout Parsing} performs a recursive division of the webpage screenshot by identifying hidden dividing lines through block projection, generating a layout division diagram and parsing the screenshot into a UI layout tree. Finally, during the \textbf{Layout-Guided Code Fusion} phase, the screenshot is divided into atomic image blocks based on location information from UI layout tree, and code snippets are sequentially generated for each image block with the help of Multimodal Large Language Models (MLLMs). These code snippets are then fused into the final synthesized code with the guidance of the UI layout tree.

\subsection{Module A: Element Relation \shuqing{Constuction}}
The goal of this module is to address the challenge of grouping UI elements that exhibit similar structures across a webpage. Through empirical analysis, we have observed that modern web designs frequently incorporate repetitive regions with analogous structures.
If these repetitive regions are not grouped and merged, directly participating in segmentation in UI Layout Parsing could lead to four potential issues: 1) excessive splitting of the layout, 2) generation of numerous redundant image blocks, 3) inconsistent code generated from redundant blocks, and 4) increased API calls per task. Hence, we propose two methods, Element Relation Mining and UI Group Search, to solve the problem of grouping repetitive regions with similar structures from the perspective of element spatial relationships.

\normalem
\vspace{-1em}
\begin{algorithm}[h]
    \SetAlgoLined
    \footnotesize
    \SetKwInOut{Input}{Input}
    \SetKwInOut{Output}{Output}
    \SetKwProg{Fn}{Function}{:}{}
    
    \Input{List of boxes $boxes$}
    \Output{List of groups $groups$}
    
    \Fn{ \textnormal{SearchUIGroups($boxes$)}}{
        $visited \gets \varnothing$\;
        $groups \gets \varnothing$\;
        
        \While{$\lvert visited \rvert  < \lvert boxes \rvert$}
        {
            $start\_bbox \gets$ ChooseRandomBBox($boxes$)\;
            $visited \gets$ Append($visited$, $start\_bbox$)\;
            \add{\codecomment{// (1) Expand group via BFS for valid nodes.}}
            
            $group \gets$ \textnormal{ExpandGroup}($start\_bbox$, $boxes$, $visited$)\;
            $groups \gets$ Append($groups$, $group$)\;
        }
        \Return{$groups$}\;
    }
    
    \vspace{0.5em} 
    
    \Fn{\textnormal{ExpandGroup}($start\_bbox, boxes, visited$)}{
        
        $group \gets start\_bbox$\;
        $queue \gets start\_bbox$\;
        
        \While{$queue \neq \varnothing$}{
            
            $current\_bbox \gets$ Pop($queue$)\;

            \add{\codecomment{// (2) Get neighbors from pre-built relation graph (Sec 3.3.1)}}
            
            $neighbors \gets$ GetBBoxNeighbors($current\_bbox$)\;

            \add{\codecomment{// (3) Traverse neighbors by alignment/spacing rules.}}
            
            \For{$neighbor$ in $neighbors$}{
                \If{$neighbor \notin visited$ \textbf{and} CanAddToGroup($neighbor$, $group$)}{
                
                    $group \gets$ Append($group$, $neighbor$)\;
                    $queue \gets$ Append($queue$, $neighbor$)\;
                    $visited \gets$ Append($visited$, $neighbor$)\;
                }
            }
        }
        \Return{$group$}\;
    }
    
\caption{UI Group Search}
\label{algo:method_A_element_relation_infer}
\end{algorithm}
\vspace{-2em}
\ULforem

\subsubsection{Element Relation Mining}
\shuqing{\tool first applies the UI element detection algorithm (UIED) \cite{UIED} to extract the bounding boxes of all text and non-text elements from the webpage image. \add{Additionally, to ensure that the obtained bounding boxes do not overlap, we sort the bounding boxes by row\_min and column\_min, and calculate four values—inter\_area, iou, ioa, and iob—between adjacent bounding boxes. Based on these values, we determine the type of overlap: if it’s an inclusion relationship, the largest bounding box is retained; if it’s an intersection relationship, the smallest enclosing rectangle is calculated.} Once the bounding boxes are obtained, we observe that each element has varying numbers of neighboring elements in different directions (i.e., above, below, left, and right).} \shuqing{For example,} in \autoref{fig:method_layoutcoder} A-a, \shuqing{element 4’s direct upper neighbors are elements 1 and 2, while its direct left neighbor is element 3. These neighboring relationships can be represented as edges in a graph, where UI elements function as nodes. To mine these spatial relationships, we traverse all the bounding boxes and construct a graph that encodes the connections between each element and its direct neighbors. This relationship graph forms the foundation for subsequent layout parsing and ensures that structurally similar regions are appropriately grouped, thereby reducing redundancy and improving the consistency of the generated code.}

\subsubsection{UI Group Search}
With the relationship graph generated from Element Relation Mining, we proceed to identify valid groupings. Prior studies indicate that UI elements with similar structures commonly exhibit alignment and consistent spacing patterns \cite{PsychologicallyUIGrouping,ScreenRecognition}.
Research in UI grouping has also explored spatial clustering methods, such as DBSCAN \cite{DBSCAN}, and principles from Gestalt psychology for effective pairing and grouping \cite{GestaltPrinciples}.
\add{For our grouping process, alignment and spacing criteria are defined as follows: 1) Alignment: Both text and non-text elements are aligned to the left, ensuring a consistent left-to-right flow within the layout; 2) Spacing: The vertical and horizontal spacing between elements is consistent. However, for groups with fewer than or equal to two elements in either the vertical or horizontal direction, spacing is not a strict requirement.}
Instead of using complex semantic grouping methods, we employ a
simpler heuristic based on alignment and spacing\delete{, leveraging the prevalent left-top alignment characteristic of text and non-text elements}. All elements are considered potential grouping candidates. Starting with a randomly selected element, we traverse its neighboring elements in the relationship graph, identifying those that meet alignment and spacing criteria. Using a breadth-first search approach, we iteratively expand the current group by adding eligible neighboring elements until the queue is empty. This process repeats for all unvisited elements until every element has been processed.
Algorithm \ref{algo:method_A_element_relation_infer} describes this grouping process, while \autoref{fig:method_layoutcoder} A-b illustrates an example where six elements are grouped based on alignment and equal spacing conditions, forming a structured grid layout.

\subsection{Module B: UI Layout Parsing}

\begin{figure}[h]
    \vspace{-1em}
    \centering
    \begin{subfigure}{0.3\textwidth}
        \centering
        \includegraphics[width=\linewidth]{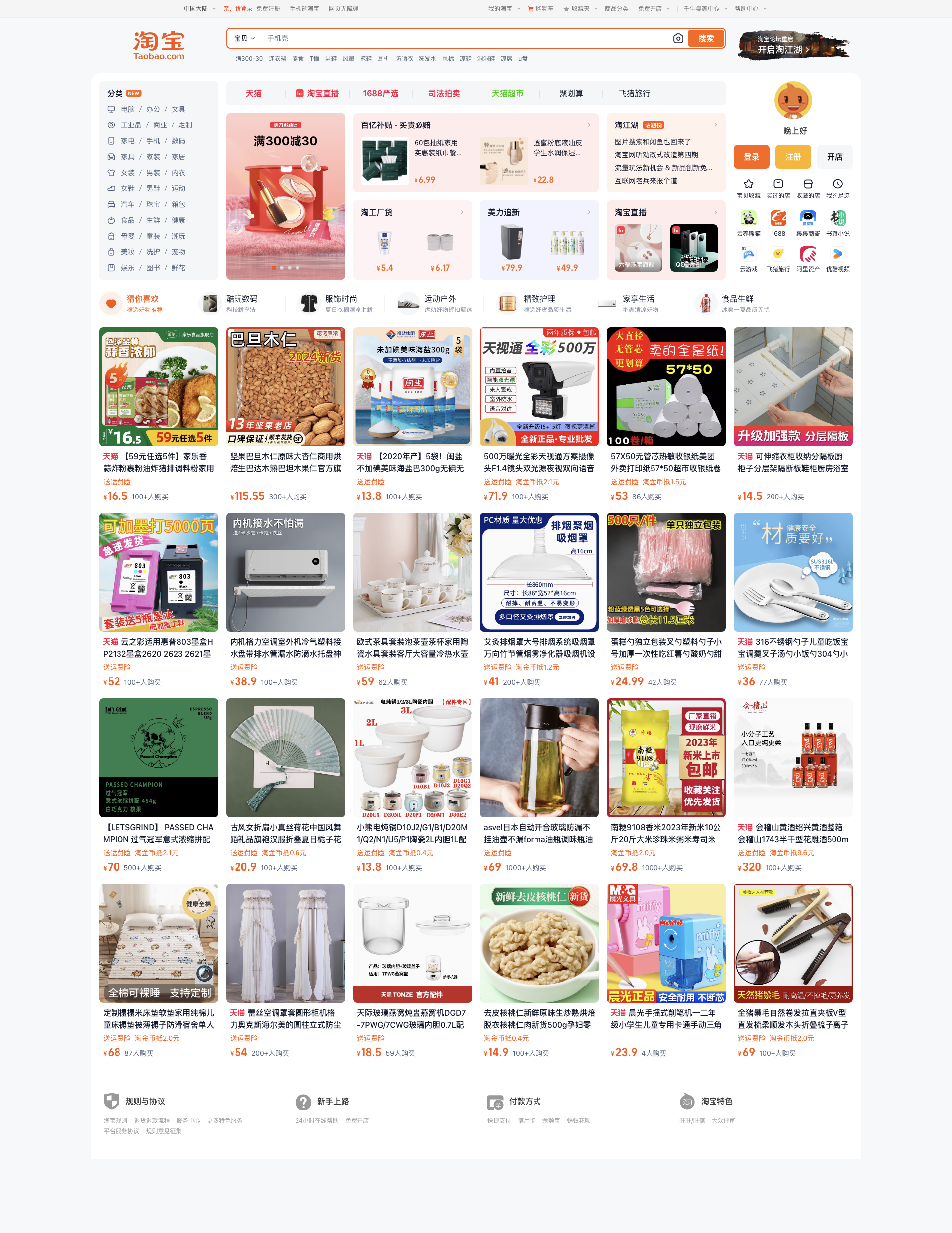}
        \caption{Reference}
        \label{fig:method_ref_taobao}
    \end{subfigure}
    \begin{subfigure}{0.3\textwidth}
        \centering
        \includegraphics[width=\linewidth]{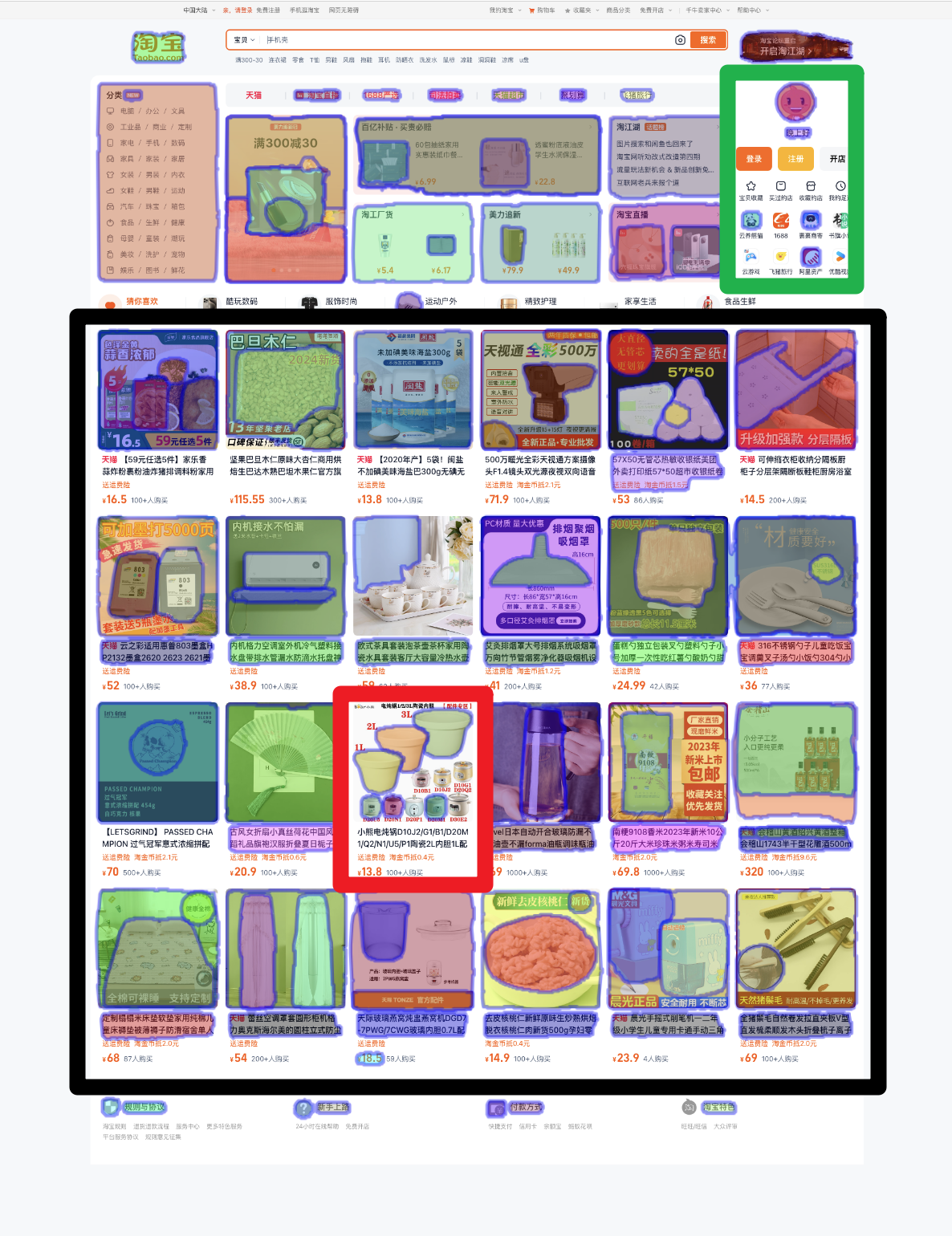}
        \caption{SAM}
        \label{fig:method_sam_taobao}
    \end{subfigure}
    \begin{subfigure}{0.3\textwidth}
        \centering
        \includegraphics[width=\linewidth]{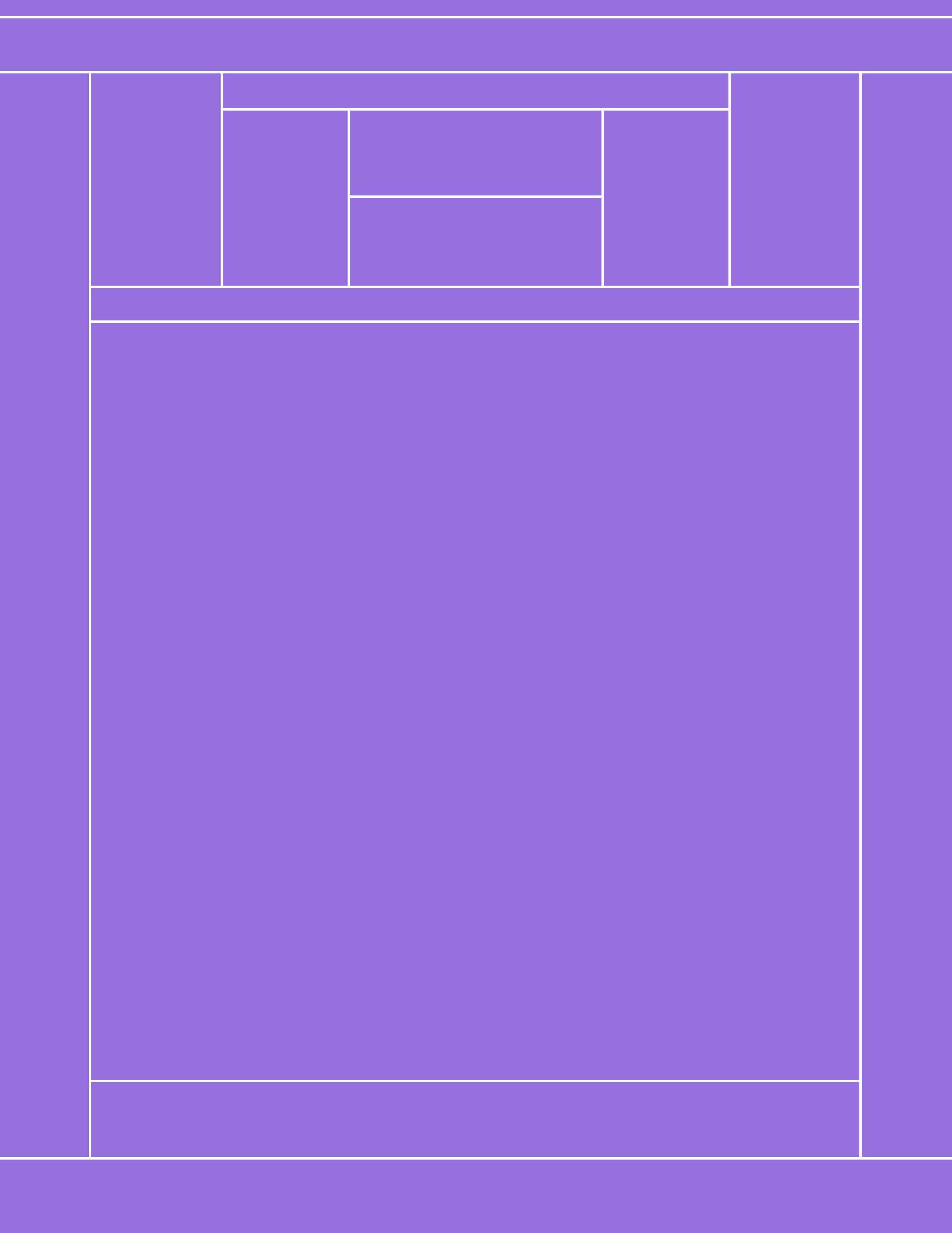}
        \caption{LayoutCoder}
        \label{fig:method_layoutcoder_taobao}
    \end{subfigure}
    \vspace{-1em}
    \caption{\yun{Illustration}
    of segmentation by SAM(b) and LayoutCoder(c).}
    \vspace{-1em}
    \label{fig:method_seg_comp}
\end{figure}

This module addresses the challenge of accurately segmenting the layout of UI images. Before designing this module, we conducted some preliminary studies on layout segmentation using existing image segmentation techniques, such as SAM.
\add{Here we conduct a case study based on taobao.com, highlighting the limitations of SAM in UI images. As illustrated in Figure \ref{fig:method_seg_comp}(b), our analysis identified several challenges:
1) Difficulty in preserving small UI components, referring to the green box in Figure \ref{fig:method_seg_comp}(b); 2) Object sensitivity in UI images, referring to the red box in Figure \ref{fig:method_seg_comp}(b); 3) Failure to cluster repeated UI components of identical types, referring to the black box in Figure \ref{fig:method_seg_comp}(b). 
To alleviate the above limitations 
, we propose three methods within the \tool framework,}
(1) 2D Block Projection, (2) Recursive Division, and (3) UI Layout Tree Construction, to segment UI images and capture layout information generated during the segmentation process.

\subsubsection{2D Block Projection}
\begin{figure}[h]
    \centering
    \vspace{-1em}
    \begin{subfigure}{0.45\textwidth}
        \centering
        \includegraphics[width=\linewidth]{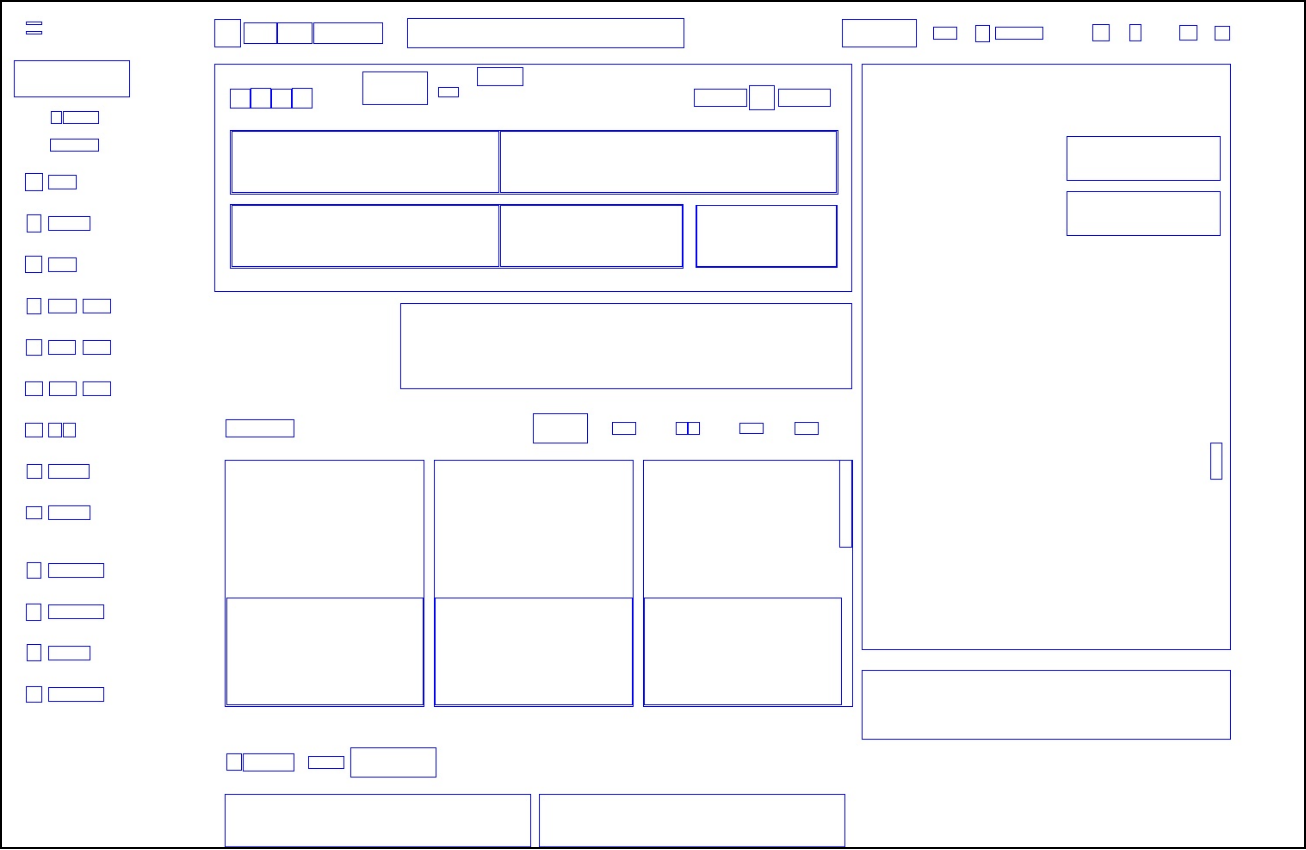}
        \caption{UIED}
        \label{fig:method_2_projection_uied}
    \end{subfigure}
    \hspace{10mm}
    \begin{subfigure}{0.45\textwidth}
        \centering
        \includegraphics[width=0.93\linewidth]{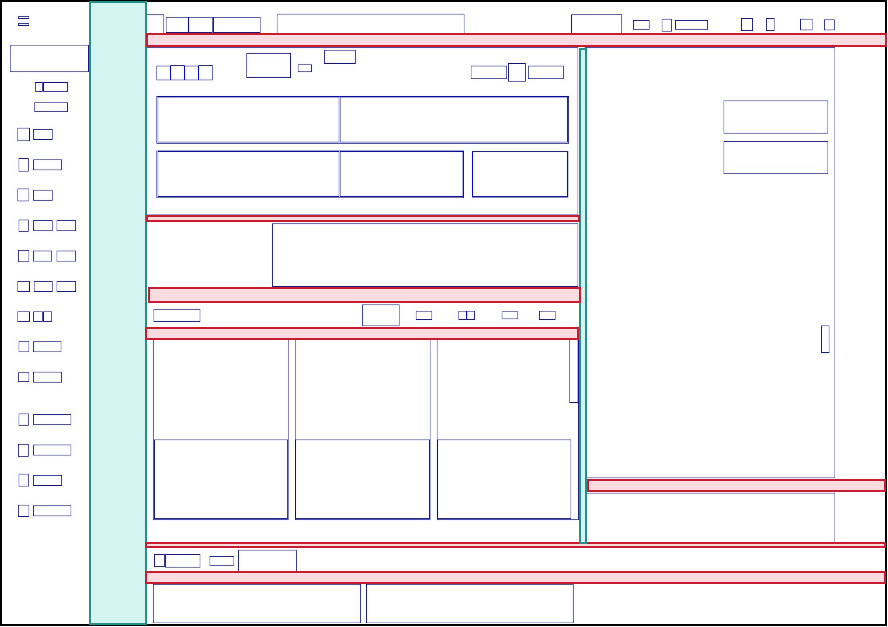}
        \caption{2D Block Projection}
        \label{fig:method_2_projection_projection}
    \end{subfigure}
    \vspace{-1em}
    \caption{An example of UIED and 2D Block Projection processing a webpage image.}
    \vspace{-1em}
    \label{fig:method_2_projection}
\end{figure}

As shown in \autoref{fig:method_layoutcoder} B-a, projecting element blocks generates projection intervals that correspond to the positions of dividing lines. The projection process can merge overlapping elements in the same direction, facilitating the identification of blank regions and preventing element truncation during segmentation. For instance, as illustrated in \autoref{fig:method_2_projection_projection} (a \shuqing{\textit{ctrip.com}} homepage example), the bounding boxes extracted by UIED are represented in rectangle boxes, while the blue and red areas visualize the block projection process, indicating potential layout divider lines' positions. After grouping elements in Element Relation Constuction, we project the blocks horizontally and vertically, collecting the projection intervals as segmentation targets for Recusive Division.

\subsubsection{Recusive Division}
Using the horizontal and vertical intervals collected in 2D Block Projection, we sort them in descending order of distance. The rationale for this approach lies in the proximity principle of Gestalt design psychology \cite{GestaltPrinciples}: elements that are closer together are perceived as related, while those further apart are seen as separated. After sorting the intervals, we first select the projection region with the largest interval and perform a midpoint division, following the proximity principle. This recursive process splits the layout into multiple regions. Each region is treated as an initial block and subjected to further projection, sorting, and division until no further divisions are possible. \autoref{fig:method_layoutcoder} B-b shows an example where the midpoint of the interval is selected as the dividing point, and the corresponding line becomes the target division line. The effectiveness of this method can be referenced in \autoref{fig:method_layoutcoder_taobao}.

\subsubsection{UI Layout Tree Construction}
We observed that the recursive segmentation process corresponds to a tree structure, where the resulting segments are nodes of the tree. This tree structure shares similarities with the \shuqing{HTML DOM tree}. Therefore, during the segmentation process in Recursive Division, we store the layout tree $T_{layout}$. This tree facilitates accurate restoration of the original layout and supports the subsequent code generation and fusion process. As shown in \autoref{fig:method_layoutcoder} B-c, we designed an attribute table that records each node's type, position, child nodes, code block, and spatial ratio. Nodes are classified into two main types: container nodes (row and column containers representing layout relationships between regions) and atomic nodes (indivisible regions). The position attribute records each region's bounding box, and the code block attribute reserves space for storing code snippets for each region.

\subsection{Module C: Layout-Guided Code Fusion}
This module focuses on code generation and fusion. 
We \shuqing{observe} that DCGen uses MLLMs for code fusion, but this approach is limited by (1) the ability of text to express layout is constrained, and (2) MLLMs have limitations in code generation for complex layouts. 
To address this, we divide the overall code generation task into two sub-tasks. The first sub-task deals with code generation for atomic regions within the leaf nodes of UI layout tree, and the second sub-task focuses on handling the layout relationship between code snippets within leaf nodes of UI layout tree. 
 For each indivisible atomic region(i.e., each leaf node of UI layout tree), we use MLLMs to generate code snippets. After finishing \shuqing{the code snippet generation of each leaf node, we convert the UI layout tree without code into the UI layout tree with code attached to the} leaf nodes. Finally, by \shuqing{the guidance of the layout relationship within the UI layout tree, code snippets within leaf nodes are fused into the} final code.
  
 This approach leverages the strengths of MLLMs in generating UI code for simple web pages while mitigating their limitations in code generation \shuqing{for web pages with complex layouts}.

\subsubsection{Code Snippet Generation}

\normalem
\vspace{-1em}
\begin{algorithm}[h]
    \SetAlgoLined
    \footnotesize
    \SetKwInOut{Input}{Input}
    \SetKwInOut{Output}{Output}
    \SetKwProg{Fn}{Function}{:}{}

    \Input{JSON layout structure $structure$}
    \Output{HTML and CSS representation $html$}
    
    \Fn{\textnormal{CodeFusion}($root\_node$)}{
        \add{\codecomment{// (1) Traverse from the root node.}} \\
        \Return \textnormal{ProcessNode}($root\_node$, \textnormal{True})\;
    }
    
    \vspace{0.5em} 
    
    \Fn{\textnormal{ProcessNode}($node, is\_root$)}{
        Initialize string $html$ as empty\;
        
        \uIf{$node[\textnormal{"type"}] = \textnormal{'row'}$}{
            \add{\codecomment{// (2) Singe Div Generation.}} \\
            $html \gets$ \textnormal{Append}($html$, ``<div class="row'')\;
            $html \gets$ \textnormal{Append}($html$, `` root"'' \textnormal{if} $is\_root$ \textnormal{else} ``"'')\;
            $html \gets$ \textnormal{Append}($html$, ``" style="flex: '', $node.get(\textnormal{"portion"}, 1)$, ``;">\textbackslash n'')\;
            {\add{\codecomment{// (3) Multi-Layer Div Generation}}} \\
            \ForEach{$child$ in $node.value$}{
                $html \gets$ \textnormal{Append}($html$, \textnormal{ProcessNode}($child$))\;
            }
            $html \gets$ \textnormal{Append}($html$, ``</div>\textbackslash n'')\;
        }
        \uElseIf{$node[\textnormal{"type"}] = \textnormal{'column'}$}{
            \add{\codecomment{// (2) Singe Div Generation.}} \\
            $html \gets$ \textnormal{Append}($html$, ``<div class="column'')\;
            $html \gets$ \textnormal{Append}($html$, `` root"'' \textnormal{if} $is\_root$ \textnormal{else} ``"'')\;
            $html \gets$ \textnormal{Append}($html$, ``" style="flex: '', $node.get(\textnormal{"portion"}, 1)$, ``;">\textbackslash n'')\;
            {\add{\codecomment{// (3) Multi-Layer Div Generation}}} \\
            \ForEach{$child$ in $node.value$}{
                $html \gets$ \textnormal{Append}($html$, \textnormal{ProcessNode}($child$))\;
            }
            $html \gets$ \textnormal{Append}($html$, ``</div>\textbackslash n'')\;
        }
        {\add{\codecomment{// (4) Connect the div element for the leaf node of code snippet}}} \\
        \ElseIf{$node[\textnormal{"type"}] = \textnormal{'atomic'}$}{
            $html \gets$ \textnormal{Append}($html$, ``<div class="atomic" style="flex: '', $node[\textnormal{"portion"}]$, ``;">'', $node[\textnormal{"code"}]$, ``</div>\textbackslash n'')\;
        }
        \Return{$html$}\;
    }
    
\caption{Code Fusion}
\label{algo:code_fusion}
\end{algorithm}
\vspace{-1em}
\ULforem

This method handles the first sub-task: code generation for indivisible atomic regions. In the UI Layout Parsing Module, we mentioned that the layout tree's attribute table includes a "position" attribute that stores each atomic region's bounding box. A webpage image contains multiple atomic regions, and by locating their bounding boxes, we can crop out the image blocks to be processed. The prompt for code snippet generation, combined with the image blocks, is batched into the MLLM, generating code snippets $C^i_{part}$ for each block. \autoref{fig:method_layoutcoder} C-a illustrates the process. The prompt for code snippet generation differs from that for complete HTML code generation. Specifically, we impose the following core requirements when designing the prompt for MLLM:
1) The generated code must start with a <div> tag and end with a </div> tag;
2) The code should not impose fixed width and height settings;
3) The image proportions should be preserved;
4) The margin and padding should be set to 0;
5) Make sure the generated code doesn't conflict with outer divs in layout and style.The first requirement ensures that the code generated represents a code snippet rather than a complete HTML code. The second and third requirement avoids MLLM's tendency to set fixed sizes and ignore the proportion when restoring images. \shuqing{The fourth and fifth ensure smooth integration into the layout code, which forms} the basis of div decoupling that ensures no conflict between inner and outer divs.

\add{After finishing the code snippet generation of all atomic regions, code snippets are attached to leaf nodes of the UI layout tree. The initial UI layout tree without code is transformed into a UI layout tree with code as shown in \autoref{fig:method_layoutcoder} C-b.}

\subsubsection{Code Fusion}

This method handles the second sub-task: leverage the layout relationship between leaf nodes of the UI layout tree to fuse code snippets. 
\add{In Algorithm \ref{algo:code_fusion}, 
\textbf{(1) Traversal:}
We first traverse the nodes of the UI Layout tree layer by layer, handling non-leaf and leaf nodes separately. Non-leaf nodes are container nodes of type row or column, while leaf nodes are atomic nodes of type atomic. 
\textbf{(2) Single Div Generation:}
\wxcl{Then,} each node in the UI Layout Tree eventually generates a corresponding div. Since there is a decoupling relationship between inner and outer div elements in HTML, there is also a decoupling relationship between nodes.
Based on this property, we control the spatial arrangement of nodes within different container nodes through associated styles using class, and control the spatial proportion of the corresponding div through inline styles. \textbf{(3) Multi-Layer Div Generation:} Layer by layer, the nodes of the UI Layout Tree are converted into multi-layer nested divs that can represent the UI Layout, with the leaf nodes containing local code snippets placed under the corresponding `div` of their container nodes. When the depth-first traversal is complete, the UI Layout Tree with code snippets is transformed into the final code.}

\section{Experimental Setup}
\subsection{Research Questions}\label{sec:setup_rq}

In this section, we evaluate the effectiveness of \tool by comparing it with the state-of-the-art approaches.
We mainly focus on the following Research Questions (RQs):

\textbf{RQ1:} How effective is \tool compared with existing UI2Code methods?

\textbf{RQ2:} What is the impact of key components on the performance of \tool for UI2Code?

\textbf{RQ3:} 
What is the influence of the complexity of real-world websites on the performance of \tool?

\textbf{RQ4:} What are the results of user studies and human evaluations?

\delete{To answer \textbf{RQ1}, we compare the three representative baselines for UI2Code, including pix2code, WebSight VLM-8B, and GPT-4o.}
\add{To answer \textbf{RQ1}, we compare the five representative models for UI2Code, including pix2code, WebSight VLM-8B, DeepSeek VL-7B, GPT-4o and Claude 3.5 Sonnet.}
To answer \textbf{RQ2}, we explore the impact of three key components of \tool
to conduct ablation studies, including UI Grouping, Gap Sorting, Custom Prompt of code snippet generation(Prompt).
This allows us to evaluate the independent contributions of each component to the overall framework. 
To answer \textbf{RQ3}, 
we select four representative metrics to measure website complexity: the depth of the DOM tree, the number of DOM tags, the number of tokens in the HTML code, and the aspect ratio of images. 
We evaluate the impact of increasing web page complexity on \tool. 
To answer \textbf{RQ4}, 
\delete{we conduct a human evaluation to qualitatively evaluate the web pages generated by the best-performing baseline GPT-4o and \tool.} \add{we conduct a human evaluation to qualitatively evaluate the web pages generated by the best-performing baseline Claude 3.5 Sonnet(Self-Refine) and \tool.} In total, three participants with experience in software development or UI design are invited to take part in the evaluation.

\subsection{Datasets}\label{sec:setup_dataset}
In this work, we evaluate \tool on two real-world UI-to-Code datasets: Design2Code \cite{design2code_dataset} and Snap2Code. 
Specifically,
\textbf{(1) Design2Code:} Design2Code \cite{design2code_dataset} is an open-source, high-quality dataset composed of 484 real-world web pages from the C4 validation set, primarily used to test MLLMs in converting visual designs to code. The collection of 250 website samples from Design2Code dataset contains an average of 158 tags, where each HTML code file has an average length of 625 tokens and each web page image has an average aspect ratio of 1.1338, as shown in \autoref{tab:dataset_stat}.
It is important to note that in the Design2Code dataset, all images on the web pages have been replaced with placeholder images.
\add{\textbf{(2)Snap2Code:} To evaluate the performance on web pages with more complex layouts and components, we also created a new dataset called Snap2Code. Furthermore, to comprehensively assess the performance of our framework on unseen datasets, the Snap2Code dataset is divided into two parts: \textit{Seen} and \textit{Unseen}. The \textit{Seen} portion is compiled from the top 500 globally visited websites, as ranked by Moz\footnote{https://moz.com/top500}, while the \textit{Unseen} portion consists of 100 newly registered web domains from the past month (as of January 3, 2025), sourced via Shreshtait\footnote{https://shreshtait.com/newly-registered-domains/nrd-1m}. We use the Selenium framework\footnote{https://www.selenium.dev/} and GeckoDriver\footnote{https://github.com/mozilla/geckodriver} to ensure high fidelity and accuracy in capturing webpage data. During the collection process, any web pages that were blocked, denied access, blank, or failed to load were excluded to maintain the quality of the dataset. For the \textit{Unseen} portion of the dataset, we checked the first archive time of the collected webpages using the API of The Wayback Machine\footnote{https://web.archive.org/} to ensure that all webpages were created after August 7, 2024, making them unseen by the selected version of MLLM (e.g., GPT-4o-2024-08-06, Claude-3-5-Sonnet-202406).}

\begin{figure}[h]
    \centering
    \vspace{-1em}
    \begin{subfigure}{0.3\textwidth}
        \centering
        \includegraphics[width=\linewidth]{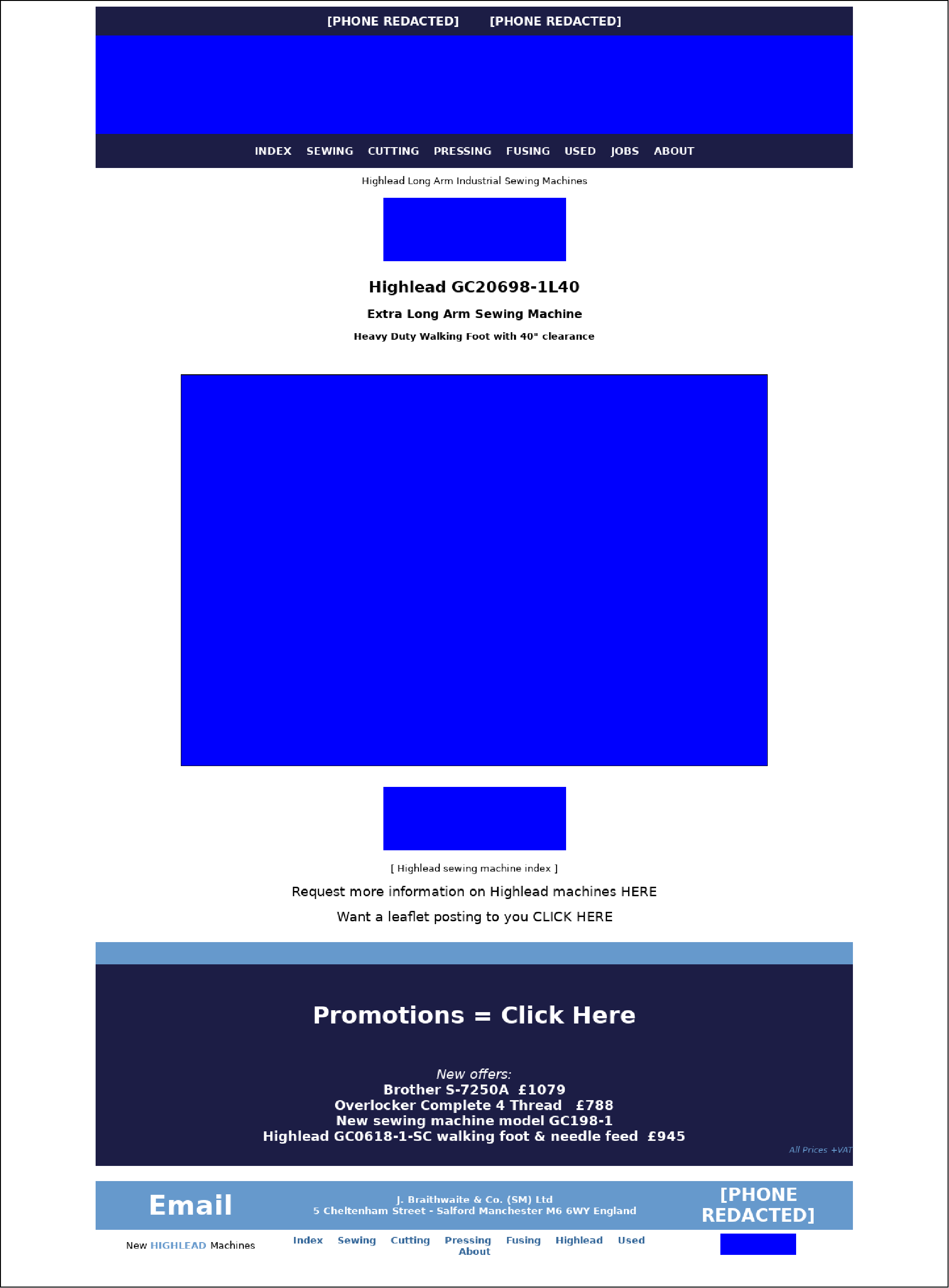}
        \caption{Design2Code}
        \label{fig:dataset_4241}
    \end{subfigure}
    \hspace{10mm}
    \begin{subfigure}{0.3\textwidth}
        \centering
        \includegraphics[width=0.9\linewidth]{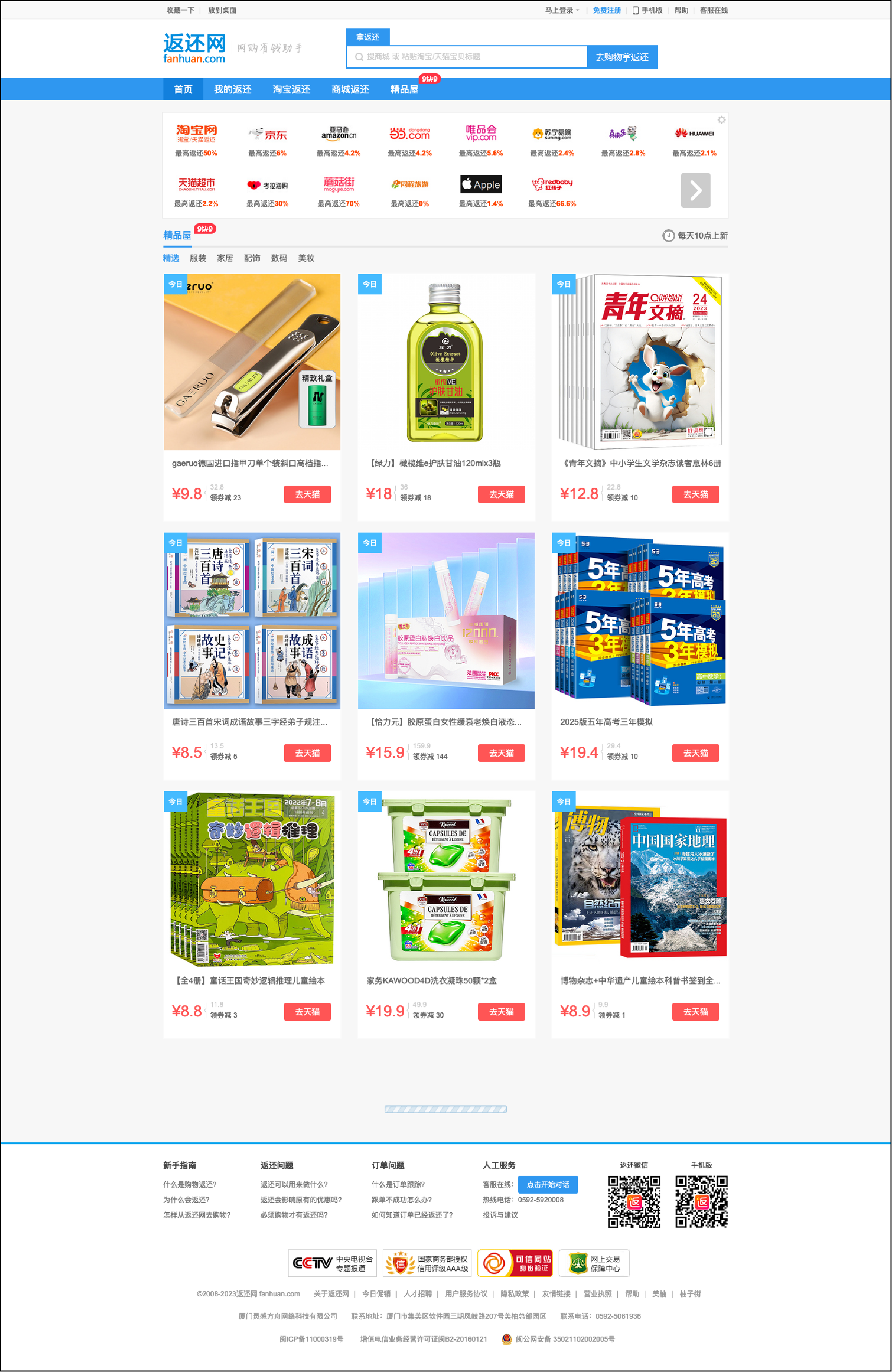}
        \caption{Snap2Code}
        \label{fig:dataset_fanhuan}
    \end{subfigure}
    \vspace{-1em}
    \caption{Two examples from Design2Code(a) and Snap2Code(b).}
    \vspace{-1em}
    \label{fig:dataset_example}
\end{figure}

\add{\textbf{Comparison of Dataset Complexity.} \autoref{tab:dataset_cmp} illustrates the differences between the Design2Code and Snap2Code datasets in terms of
scope, complexity, and applicability. \autoref{tab:dataset_stat} mainly shows the complexity of the image and text
data in Design2Code and Snap2Code from a data statistics perspective. Upon observation, we find that Snap2Code(Seen) exceeds Design2Code by an order of magnitude in the median values for DOM Breadth, Tag Count, and Length. In addition, the median value for the Aspect Ratio metric of images in Snap2Code (Seen) is twice that of Design2Code. This metric poses a challenge
for methods aimed at understanding longer images and generating longer code.
Additionally, Snap2Code retains the original images within the web pages, presenting a more realistic and challenging scenario for UI2Code task, as shown in \autoref{fig:dataset_example}.}

\add{\textbf{Clarification of Snap2Code(Seen) Dataset Size Inconsistency. } Although Snap2Code(Seen) is compiled from the top 500 globally visited websites, some samples were excluded due to issues encountered during web crawling, such as being blocked, denied access, displaying blank pages, or failing to load. Additionally, considering that the GPT-4o and Claude 3.5 Sonnet models are quite expensive, we set the dataset size to 250, following the methods from previous works \cite{REMAUI, DCGen, DeclarUI}, which led to this Snap2Code(Seen) dataset size inconsistency. }

\begin{table}[t]
\vspace{-1em}
\caption{The comparison of Snap2Code with Design2Code: scope, complexity, and applicability.}
\vspace{-1em}
\label{tab:dataset_cmp}
\resizebox{\textwidth}{!}{%
\begin{tabular}{lllll}
\hline
                  & Scope                                                                   & Sample Size & Complexity                                    & Applicability  \\ \hline
Design2Code       & 484 real-world web pages from the C4 validation set                     & 250         & Simple(All images replaced with placeholders) & Web UI-to-code \\
Snap2Code(Seen)   & the top 500 globally visited websites                                   & 250         & Complex(Original images retained)           & Web UI-to-code \\
Snap2Code(Unseen) & newly registered web domains from the past month as of January 3, 2025  & 100         & Moderate(Original images retained)           & Web UI-to-code \\ \hline
\end{tabular}%
}
\vspace{-1em}
\end{table}

\begin{table}[t]
\vspace{-1em}
\caption{The statistics of both Design2Code and Snap2Code.}
\vspace{-1em}
\label{tab:dataset_stat}
\resizebox{\textwidth}{!}{%
\begin{tabular}{clrrrrrr}
\hline
\multicolumn{1}{l}{\textbf{Dataset}}         & \textbf{Metric} & \multicolumn{1}{l}{\textbf{DOM Depth}} & \multicolumn{1}{l}{\textbf{DOM Breadth}} & \multicolumn{1}{l}{\textbf{Tag Count}} & \multicolumn{1}{l}{\textbf{Unique Tags}} & \multicolumn{1}{l}{\textbf{Length}} & \multicolumn{1}{l}{\textbf{Aspect Ratio}} \\ \hline
                                             & \textbf{Mean}   & 13                                     & 80                                       & 158                                    & 23                                       & 625                                 & 1.13                                    \\
                                             & \textbf{Std}    & 5                                      & 51                                       & 100                                    & 6                                        & 338                                 & 0.50                                    \\
                                             & \textbf{Min}    & 5                                      & 6                                        & 13                                     & 10                                       & 34                                  & 0.56                                    \\
                                             & \textbf{Median}   & 12                                     & {\color[HTML]{000000} 67}                & {\color[HTML]{000000} 134}             & 23                                       & {\color[HTML]{000000} 571}          & {\color[HTML]{000000} 1.04}             \\
\multirow{-8}{*}{\textbf{Design2Code}}       & \textbf{Max}    & 26                                     & {\color[HTML]{000000} 260}               & {\color[HTML]{000000} 486}             & 46                                       & {\color[HTML]{000000} 1679}         & {\color[HTML]{000000} 2.60}             \\ \hline
                                             & \textbf{Mean}   & 20                                     & 702                                      & 1468                                   & 37                                       & 4379                                & 2.74                                    \\
                                             & \textbf{Std}    & 7                                      & 772                                      & 1580                                   & 11                                       & 5029                                & 1.53                                   \\
                                             & \textbf{Min}    & 6                                      & 6                                        & 14                                     & 11                                       & 68                                  & 0.56                                   \\
                                             & \textbf{Median}   & 20                                     & {\color[HTML]{000000} 524}               & {\color[HTML]{000000} 1118}            & 37                                       & {\color[HTML]{000000} 3107}         & {\color[HTML]{000000} 2.79}             \\
\multirow{-8}{*}{\textbf{Snap2Code(Seen)}}   & \textbf{Max}    & 43                                     & {\color[HTML]{000000} 8787}              & {\color[HTML]{000000} 15671}           & 94                                       & {\color[HTML]{000000} 53682}        & {\color[HTML]{000000} 5.98}             \\ \hline
                                             & \textbf{Mean}   & 18                                     & 168                                      & 332                                    & 30                                       & 1077                                & 1.42                                   \\
                                             & \textbf{Std}    & 7                                      & 164                                      & 326                                    & 11                                       & 1284                                & 0.98                                    \\
                                             & \textbf{Min}    & 7                                      & 13                                       & 25                                     & 13                                       & 44                                  & 0.59                                   \\
                                             & \textbf{Median}   & 17                                     & {\color[HTML]{000000} 116}               & {\color[HTML]{000000} 214}             & 29                                       & {\color[HTML]{000000} 585}          & {\color[HTML]{000000} 0.96}             \\
\multirow{-8}{*}{\textbf{Snap2Code(Unseen)}} & \textbf{Max}    & 45                                     & {\color[HTML]{000000} 963}               & {\color[HTML]{000000} 1888}            & 60                                       & {\color[HTML]{000000} 6701}         & {\color[HTML]{000000} 5.16}             \\ \hline
\end{tabular}%
}
\end{table}

\subsection{Baselines}\label{sec:setup_baseline}

\add{To evaluate the performance of \tool, we select five representative models (pix2code \cite{pix2code}, WebSight VLM-8B \cite{websight_dataset}, DeepSeek VL-7B, GPT-4o, and Claude 3.5 Sonnet) and three different levels of prompts (Direct, CoT, and Self-Refine) targeting closed-source models, resulting in a total of nine baselines.} The \textbf{pix2code} integrates CNN and LSTM models to generate code from images, which is one of the few representative open-source models available for UI-to-Code tasks in web applications. \textbf{WebSight VLM-8B} is an open-source VLM (vision-language model)  from Huggingface, fine-tuned on a synthetic UI-to-Code dataset, WebSight-v0.1\footnote{https://huggingface.co/datasets/HuggingFaceM4/WebSight}. 
\add{
\textbf{DeepSeek VL-7B} is an open-source Vision-Language model with broad multimodal understanding, capable of processing diverse real-world applications.
\textbf{GPT-4o}, developed by OpenAI, is a closed-source multimodal large language model (MLLM) that excels in image-to-text and UI-to-Code tasks, which serves as a key component in the Code Fusion module of the \tool framework. Similarly, \textbf{Claude 3.5 Sonnet}, developed by Anthropic, offer exceptional capabilities in code generation and visual understanding.}

\add{For GPT-4o, we set the model to version "GPT-4o-2024-08-06," while for Claude 3.5 Sonnet, the version is "Claude-3-5-Sonnet-20240620" both with a maximum token limit of 4096 and a temperature of 0. Additionally, The three different levels of prompts are based on the prompt schemes from Design2Code and DCGen. We apply them to GPT-4o and Claude 3.5 Sonnet. The specific content of the prompts will be made available in the anonymous code repository. In this paper, we will refer to Claude(D), Claude(C), and Claude(SR) as Claude 3.5 Sonnet+Direct, Claude 3.5 Sonnet+CoT, and Claude 3.5 Sonnet+Self-Refine, respectively, with similar naming for GPT-4o.
}

\subsection{Metrics}\label{sec:setup_metric}
\wxc{Following the previous work \cite{DCGen, design2code_dataset, DeclarUI}}
, we employ two 
\wxc{widely-used}
metrics: BLEU score \cite{BLEUscore} and CLIP score \cite{CLIPscore}, to evaluate the similarity between the generated webpage code and the rendered webpage focusing on textual and visual aspects.

\wxc{To evaluate the textual similarity,}
we use the \textbf{BLEU score} to evaluate the similarity between the generated code ($C_g$) and the original code ($C_o$). The BLEU score is a widely used metric in natural language processing to assess the similarity of generated text to reference text based on n-gram matching. 
\wxc{It} is computed as: $ BLEU - N = BP \times \exp\left (\sum_{n=1}^Nw_n\log{P_n}\right ) $
where $P_n$ \wxc{denotes} the n-gram precision, and $w_n$ represents the weight for each n-gram.
$N$ is the maximum n-gram order.
Brevity Penalty (BP) is applied to penalize generated texts that are significantly shorter than the reference text. If the generated text length $c$ is shorter than the reference text length $r$, $BP = \exp\left ( 1 - \frac{r}{c} \right )$; otherwise, $BP = 1$. 

\wxc{To evaluate the visual similarity,}
we use the \textbf{CLIP score} to assess the similarity between the webpage screenshot rendered from the generated code ($I_g$) and the screenshot rendered from the original code ($I_o$). CLIP score is a metric in computer vision that evaluates the visual similarity between generated and reference images. The images are converted into embedding vectors by CLIP’s image encoder, and the cosine similarity between these vectors determines their similarity in the CLIP embedding space \cite{CLIP}. In particular, we extract features using the ViT-B/32 model.

\subsection{Implementation Details}
For the execution environment, we run Python programs with Python 3.8.18 and manage dependencies using conda 24.5.0. All experiments are conducted on a Linux server (64-bit Ubuntu 20.04) equipped with two 48-core Intel Xeon Platinum 8260C CPUs @ 2.40GHz (96 logical processors), four NVIDIA A100-40GB GPUs, and 251GB RAM.

\section{Experimental Results}\label{sec:exp_result}
This section presents our experimental results and answers the four RQs outlined in Section \ref{sec:setup_rq}.

\subsection{RQ1: Comparison with Baselines}

\begin{figure}[ht]
    \vspace{-1em}
    \centering
    \begin{subfigure}{\textwidth}
        \centering
        \includegraphics[width=\linewidth]{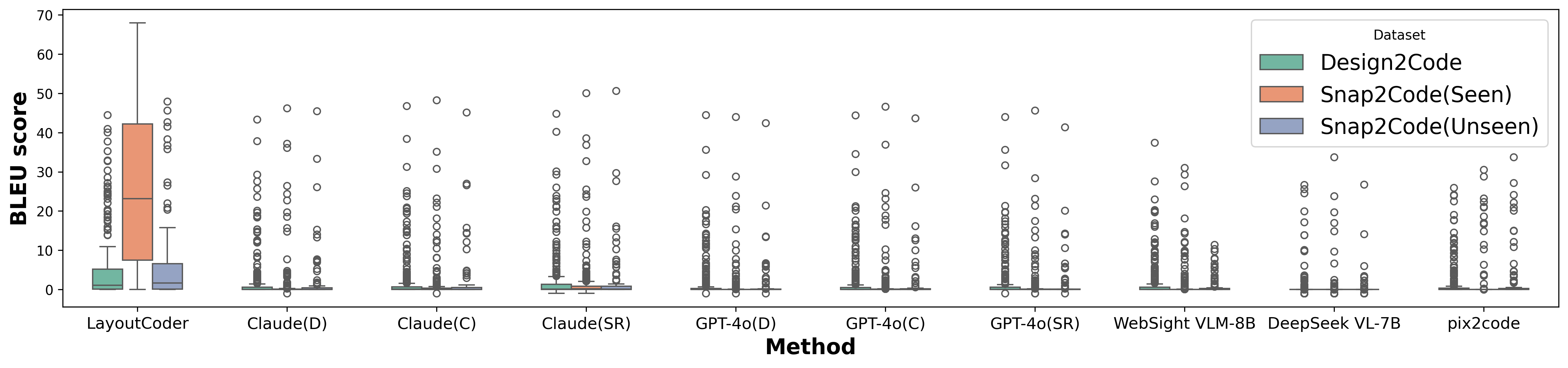}
        \vspace{-2em}
        \caption{Distribution of BLEU Scores}
        \label{fig:rq1_box_plot_bleu}
    \end{subfigure}
    \vfill
    \begin{subfigure}{\textwidth}
        \centering
        \includegraphics[width=\linewidth]{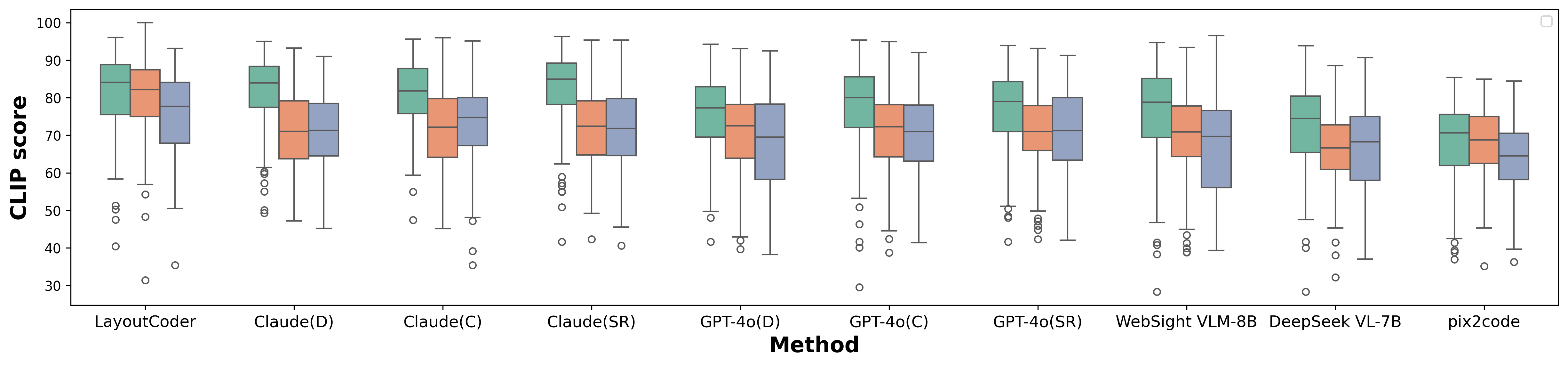}
        \vspace{-2em}
        \caption{Distribution of CLIP Scores}
        \vspace{-1em}
        \label{fig:rq1_box_plot_clip}
    \end{subfigure}
    \vspace{-1em}
    \caption{Box plots showing the distribution of BLEU and CLIP scores for \tool and baseline methods.}
    \vspace{-1em}
    \label{fig:exp_result_comparison_rq1_box_plot}
\end{figure}

\add{To answer this RQ, we compared the performance of \tool with the nine baselines.} \autoref{tab:exp_comp_rq1} shows
the experimental results of each method of BLEU
and CLIP scores.
Additionally, \autoref{fig:rq1_box_plot_bleu} and \autoref{fig:rq1_box_plot_clip} present box plots of the BLEU and CLIP scores, providing further insights into the distribution of results across the two datasets.

\noindent\textbf{Analyses:}

(1) \add{\textbf{
\tool outperforms other baselines in terms of both text similarity and visual fidelity.} 
\wxcl{When considering all the two metrics regarding the three datasets (6 combination cases altogether), \tool has the best performance in 5 out of the 6 cases. As shown in \autoref{tab:exp_comp_rq1},} it has a 24.12\% improvement in BLEU score on the Snap2Code(Seen) dataset and a 2.28\% improvement on the Design2Code dataset.
For visual consistency, \tool outperforms Claude(SR) by 8.38\% to 4.63\% in terms of CLIP score across the Seen and Unseen parts of the Snap2Code dataset. This indicates that \tool achieves better performance in both text similarity and visual fidelity. 
\wxcl{On the unseen data, while \tool shows a slight degradation compared to the seen dataset,} 
its overall performance remains optimal. 
Furthermore, the Claude model 
\wxcl{exhibits the best} CLIP score on the Design2Code dataset. 
\wxcl{It may be}
due to its advantages in visual comprehension and code generation capabilities. 
}

(2) \textbf{\tool also demonstrates advantages in the overall data distribution compared to other methods.  
} 
\add{The \autoref{fig:rq1_box_plot_bleu} illustrates that \tool's lower quartile is higher than Claude(SR)'s upper quartile in the BLEU score on the Snap2Code(Seen) dataset which indicates that over 75\% of the web pages generated by LayoutCoder surpass 75\% of those generated by Claude(SR). On the Design2Code dataset, 50\% of the web pages generated by \tool also surpass the vast majority of Claude(SR)'s, excluding some outliers. Additionally, \autoref{fig:rq1_box_plot_clip} visualizes the CLIP score distribution, which shows 25\% of the web pages generated by \tool exceeds 85.00\% in the CLIP score on the Snap2Code(Seen) dataset, while only 25\% of Claude(SR)'s approaches 80.00\%.}

Thus, \tool not only excels in average metrics but also in the overall data distribution compared to other methods.

\begin{table}[t]
\vspace{-1em}
\caption{Comparison results of LayoutCoder with three baseline methods(\%).}
\label{tab:exp_comp_rq1}
\vspace{-1em}
\resizebox{0.7\textwidth}{!}{%
\begin{tabular}{lcccccc}
\hline
\multirow{2}{*}{\textbf{Method}} & \multicolumn{2}{c}{\textbf{Design2Code}} & \multicolumn{2}{c}{\textbf{Snap2Code(Seen)}} & \multicolumn{2}{c}{\textbf{Snap2Code(Unseen)}} \\ \cline{2-7} 
                                 & \textbf{BLEU}      & \textbf{CLIP}       & \textbf{BLEU}         & \textbf{CLIP}        & \textbf{BLEU}         & \textbf{CLIP}          \\ \hline
Claude(D)                    & 2.19               & 81.90               & 1.29                  & 70.77                & 2.11                  & 71.35                  \\
Claude(C)                       & 2.31               & 81.17               & 1.39                  & 71.66                & 2.21                  & 73.38                  \\
Claude(SR)               & 2.66               & \textbf{82.74}      & 1.95                  & 71.96                & 2.45                  & 71.51                  \\ \hline
GPT-4o(D)                    & 1.86               & 75.57               & 0.97                  & 71.37                & 1.34                  & 68.11                  \\
GPT-4o(C)                       & 2.04               & 77.99               & 1.14                  & 71.22                & 1.59                  & 70.26                  \\
GPT-4o(SR)               & 1.96               & 76.70               & 1.00                  & 71.19                & 1.39                  & 70.56                  \\ \hline
pix2code(greedy)                 & 1.59               & 68.24               & 1.16                  & 68.36                & 2.47                  & 64.79                  \\
WebSight VLM-8B                  & 1.75               & 76.40               & 1.07                  & 70.41                & 1.05                  & 66.86                  \\
DeepSeek VL-7B                   & 0.69               & 72.34               & 0.45                  & 66.64                & 0.49                  & 66.65                  \\ \hline
\textbf{LayoutCoder}             & \textbf{4.93}      & 81.58               & \textbf{26.07}        & \textbf{80.35}       & \textbf{6.46}         & \textbf{76.14}         \\ \hline
\end{tabular}%
}
\vspace{-1em}
\end{table}

\finding{1}{
\add{In comparison with the best baseline, Claude(SR), \tool achieves the best performance across all evaluated performance metrics, with the improvement of \add{10.14}\% in the BLEU score and \add{3.95}\% in the CLIP score on average across all datasets.}}

\subsection{RQ2: Ablation Study}

\wxc{To answer RQ2, we explore the impact of different components of \tool, including the UI Group Search (i.e, w/o UI grouping), Recursive Division (i.e, w/o Gap Sorting), and Code Snippet Generation (i.e, w/o Prompt). \autoref{tab:exp_ab_rq2} presents the performance of \tool and its three variants on Snap2Code and Design2Code datasets.}

\subsubsection{UI Group Search}
\wxc{To investigate the impact of the UI grouping in element relation construction, we deploy one variant that removed UI grouping (i.e., w/o UI Group Search). It means that similar structural regions are not merged in a UI group.}
\wxc{As shown in \autoref{tab:exp_ab_rq2}}, 
we can find that the UI grouping can improve the performance of LayoutCoder. \add{Specifically, in the Snap2Code(Seen) dataset, the BLEU score increased by 0.55\%, the CLIP score increased by 1.51\%.}
\wxc{When UI elements are not grouped, structurally similar regions lead to inconsistencies in code generation. This lack of grouping disrupts both layout segmentation and visual rendering uniformity.}

\subsubsection{Recursive Division.
}
\wxc{To explore the impact of the recursive division in UI layout Parsing, we conduct a variant, which only uses the division without gap sorting (i.e., w/o Recursive Division), thereby disrupting the far-to-near segmentation order that adheres to the Gestalt principle of proximity. As shown in \autoref{tab:exp_ab_rq2}, we can find that the gap sorting can improve the performance of LayoutCoder.}
\add{Specifically, in the Snap2Code(Seen) dataset, the BLEU score increased by 0.19\% and the CLIP score increased by 6.30\%.} When the element shadow gaps are not sorted, unsorted gaps lead to unordered segmentation. This lack of gap sorting disrupts both layout segmentation and the accurate code generation.

\subsubsection{Simplifying the Prompt.
}
To explore the impact of div decoupling in the prompt of code snippet generation, we employ a variant that simplifies its basic form in layout-guided code fusion, thereby disrupting the compatibility of layout and style between outer divs and inner divs. As shown in \autoref{tab:exp_ab_rq2}, we can find a complete prompt can improve the performance of LayoutCoder. \add{Specifically, in the Snap2Code(Seen) dataset, the BLEU score increased by 0.03\% and the CLIP score increased by 4.83\%. }When the prompt is simplified, layout and style of inner divs conflicts with those of outer divs. This lack of the complete prompt disrupts code generation and visual rendering uniformity.

\begin{table}[t]
\vspace{-1em}
\caption{Performance of LayoutCoder and its variants on two datasets(\%).}
\label{tab:exp_ab_rq2}
\vspace{-1em}
\resizebox{0.7\textwidth}{!}{%
\begin{tabular}{lcccccc}
\hline
\textbf{Dataset}     & \multicolumn{2}{c}{\textbf{Design2Code}} & \multicolumn{2}{c}{\textbf{Snap2Code(Seen)}} & \multicolumn{2}{c}{\textbf{Snap2Code(Unseen)}} \\ \hline
\textbf{Metrics}     & \textbf{BLEU}      & \textbf{CLIP}       & \textbf{BLEU}         & \textbf{CLIP}        & \textbf{BLEU}         & \textbf{CLIP}          \\ \hline
w/o Grouping         & 4.76               & 81.21               & 25.53                 & 78.84                & 6.13                  & 74.25                  \\
w/o Gap Sorting      & 4.63               & 80.06               & 25.89                 & 74.05                & 6.19                  & 71.47                  \\
w/o Prompt           & 4.51               & 78.25               & 26.04                 & 75.51                & 6.44                  & 70.05                  \\
\textbf{LayoutCoder} & \textbf{4.93}      & \textbf{81.58}      & \textbf{26.07}        & \textbf{80.35}       & \textbf{6.46}         & \textbf{76.14}         \\ \hline
\end{tabular}%
}
\vspace{-1em}
\end{table}

\finding{2}{All three modules are essential to the \tool framework, with gap sorting (i.e., Recursive Division) and the complete prompt (i.e., Code Snippet Generation) contributing the most to overall performance.}

\subsection{RQ3: Influence of Website Complexity}

\wxc{To answer RQ3, 
\add{we conduct experiments of webpage complexity to evaluate the performance of \tool, against the leading baseline , Claude(SR).} We extract webpage code and rendered image features that effectively encapsulate the complexity of a webpage. 
Specifically, we identified and selected four representative features, including DOM depth, tag count, length and aspect ratio.}

\textbf{DOM Depth:} 
The depth of the Document Object Model (DOM) tree depth in the webpage's HTML files, representing the nesting complexity of HTML tags;
\textbf{Tag Count:} 
We quantified the structural complexity of HTML documents by counting the number of tags within each file;
\textbf{Length (tokens):} The number of tokens in the webpage's HTML code, measured using the cl100k\_base tokenizer from OpenAI's tiktoken library;
\textbf{Aspect Ratio:} The aspect ratio of the webpage images, with all images fixed at a width of 1512 pixels, reflects visual complexity.

\begin{figure} [h]
  \vspace{-1em}
  \centering
  \includegraphics[width=\linewidth]{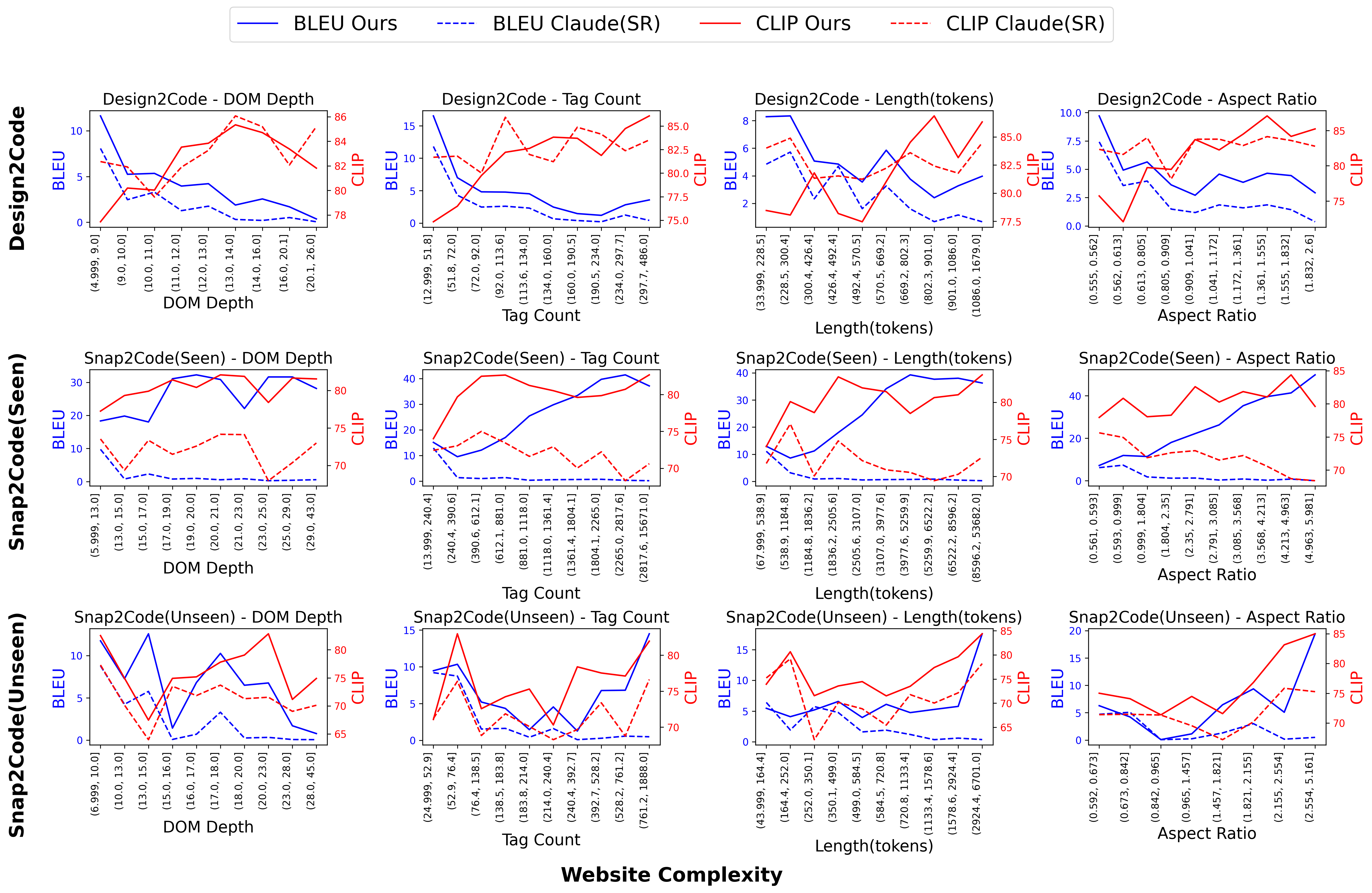}
  \vspace{-2em}
  \caption{The performance of \tool across the four complexity features for both datasets.}
  \vspace{-1em}
  \label{fig:exp_rq3_complex}
\end{figure}

Due to uneven sample densities across various complexity intervals, we implement equal-frequency binning on the x-axis to ensure a uniform distribution of data points across the range of complexities. \add{\autoref{fig:exp_rq3_complex} illustrates the performance of  \tool, compared to Claude(SR) across four distinct complexity features. Each subfigure within the figure provides a detailed visual representation of how \tool performs relative to Claude(SR) under each specific complexity feature, evaluating the effectiveness in handling varying levels of website complexity.}

\noindent \textbf{Analyses:}

(1) \add{\textbf{\tool consistently outperforms Claude(SR) across different webpage complexity dimensions in the Snap2Code dataset.} As shown in \autoref{fig:exp_rq3_complex}, we can find that
it is evident that \tool maintains a performance gap with Claude(SR) across varying levels of complexity in the Snap2Code dataset, showcasing \tool's advantage on complex web pages. 
For example, \tool achieves an 12.50\% improvement in the CLIP score on the Snap2Code(Seen) dataset for tag count in the range (2265, 2817.6] and a 15.00\% improvement for aspect ratio in the range (4.213, 4.963]. In addition, it also achieves a 15.00\% improvement in the BLEU score on the Snap2Code(Seen) dataset for length(tokens) in the range (3977.6, 5259.9].}

(2) \textbf{\tool has robust generalization on highly complex web pages.} As the website complexity increases, \tool consistently maintains a high score in both BLEU and CLIP scores. \add{For example, on the Snap2Code(Seen) dataset, the performance of \tool consistently surpasses 10.00\% in the BLEU score and 74.00\% in the CLIP score.}

\finding{3}{\add{\tool outperforms Claude(SR) across different levels of webpage complexity and demonstrates robust generalization and stability on highly complex web pages.}}

\subsection{RQ4: Human Evalution}

\add{In this section, we perform a quantitative evaluation of web pages generated by Claude 3.5 Sonnet(Self-Refine) and \tool through human evaluation. The evaluation is conducted through 
online questionnaire. Specifically, \add{four} participants with experience in software development or UI design are invited to take part in the evaluation. Each participant is asked to answer \add{120} questions, containing a total of 360 images, and rate the quality of web pages generated by Claude 3.5 Sonnet(Self-Refine) and \tool. }

\subsubsection{Survey Design}

\add{We randomly selected 50, 50 and 20 web pages from the Design2Code, Snap2Code(Seen) and Snap2Code(Unseen) datasets, respectively, for evaluation. Each question in the survey comprises a reference webpage image along with webpage images generated by the two methods , as well as the corresponding webpage code generated by both methods. Each participant will be given 120 questions. To mitigate order bias, the display order of webpage images generated by different methods will be randomized within each question. 
The quality of the generated web pages is evaluated based on four metrics: Layout Accuracy, Visual Fidelity, Content Alignment and Code Usability, using a 1-to-5 Likert scale (5 for excellent, 4 for good, 3 for acceptable, 2 for marginal, 1 for poor). At the beginning of the survey, we explained the meaning of the two evaluation metrics: Layout Accuracy measures how closely the layout structure of the generated page and the reference page; Visual Fidelity evaluates the overall visual consistency of the generated page and the reference page; Content Alignment measures the similarity between the specific content (e.g., text, images) in the generated page and the reference page; Code Usability evaluates how helpful the generated code is for developers}.

\subsubsection{Survey Result}

\begin{table}[ht]
\centering
\vspace{-1em}
\caption{Human evaluation on web pages generated by LayoutCoder and Claude(SR).}
\label{tab:human_eval}
\vspace{-1em}
\resizebox{0.6\textwidth}{!}{%
\begin{tabular}{cc|cc}
\hline
Dataset                            & Metrics           & \multicolumn{1}{l}{Claude(SR)} & \multicolumn{1}{l}{LayoutCoder} \\ \hline
\multirow{4}{*}{Design2Code}       & Layout Accuracy   & 3.25                         & \textbf{3.77}                 \\
                                   & Visual Fidelity   & 3.22                         & \textbf{3.77}                 \\
                                   & Content Alignment & 3.39                         & \textbf{4.01}                 \\
                                   & Code Usability    & 3.51                         & \textbf{4.01}                 \\ \hline
\multirow{4}{*}{Snap2Code(Seen)}   & Layout Accuracy   & 1.64                         & \textbf{4.07}                 \\
                                   & Visual Fidelity   & 1.64                         & \textbf{4.13}                 \\
                                   & Content Alignment & 1.86                         & \textbf{4.12}                 \\
                                   & Code Usability    & 1.75                         & \textbf{4.23}                 \\ \hline
\multirow{4}{*}{Snap2Code(Unseen)} & Layout Accuracy   & 1.76                         & \textbf{3.99}                 \\
                                   & Visual Fidelity   & 1.78                         & \textbf{4.00}                 \\
                                   & Content Alignment & 2.33                         & \textbf{4.03}                 \\
                                   & Code Usability    & 1.93                         & \textbf{4.11}                 \\ \hline
\end{tabular}%
}
\vspace{-1em}
\end{table}

\add{We received 480 sets of ratings from 4 participants, with each set containing 4 ratings, resulting in a total of 1920 ratings. On average, the participants spent 2 hours completing the questionnaire. The evaluation results are shown in \autoref{tab:human_eval}. We observe that LayoutCoder achieves the highest score in Code Usability on the Design2Code dataset, with a score of 4.01 and an improvement of 14.31\% over Claude(SR).} 
\add{It indicates that participants believe the generated code by LayoutCoder is useful for developers on Design2Code (good level).
In the Snap2Code(Seen) dataset, LayoutCoder significantly outperforms Claude(SR) with improvements of 152.60\% in Visual Fidelity and 122.10\% in Content Alignment, with average scores exceeding 4.00 (good level). In contrast, Claude(SR)'s performance is poor, with only 1.64 in Visual Fidelity and 1.86 in Content Alignment, indicating that the four participants are relatively dissatisfied with the results of Claude(SR) on Snap2Code(Seen) (marginal level). The results show that \textbf{LayoutCoder is better than Claude(SR) at replicating the layout and style of the reference web pages while maintaining content consistency and code usability.}}

\section{Discussion}

\subsection{ Why Does Our Model Work?}
In this paper, we provide several generation cases to demonstrate the effectiveness of \tool in the UI2Code task for web pages.

\begin{figure}[h]
    \vspace{-1em}
    \centering
    \begin{subfigure}{0.25\textwidth}
        \centering
        \includegraphics[width=\linewidth]{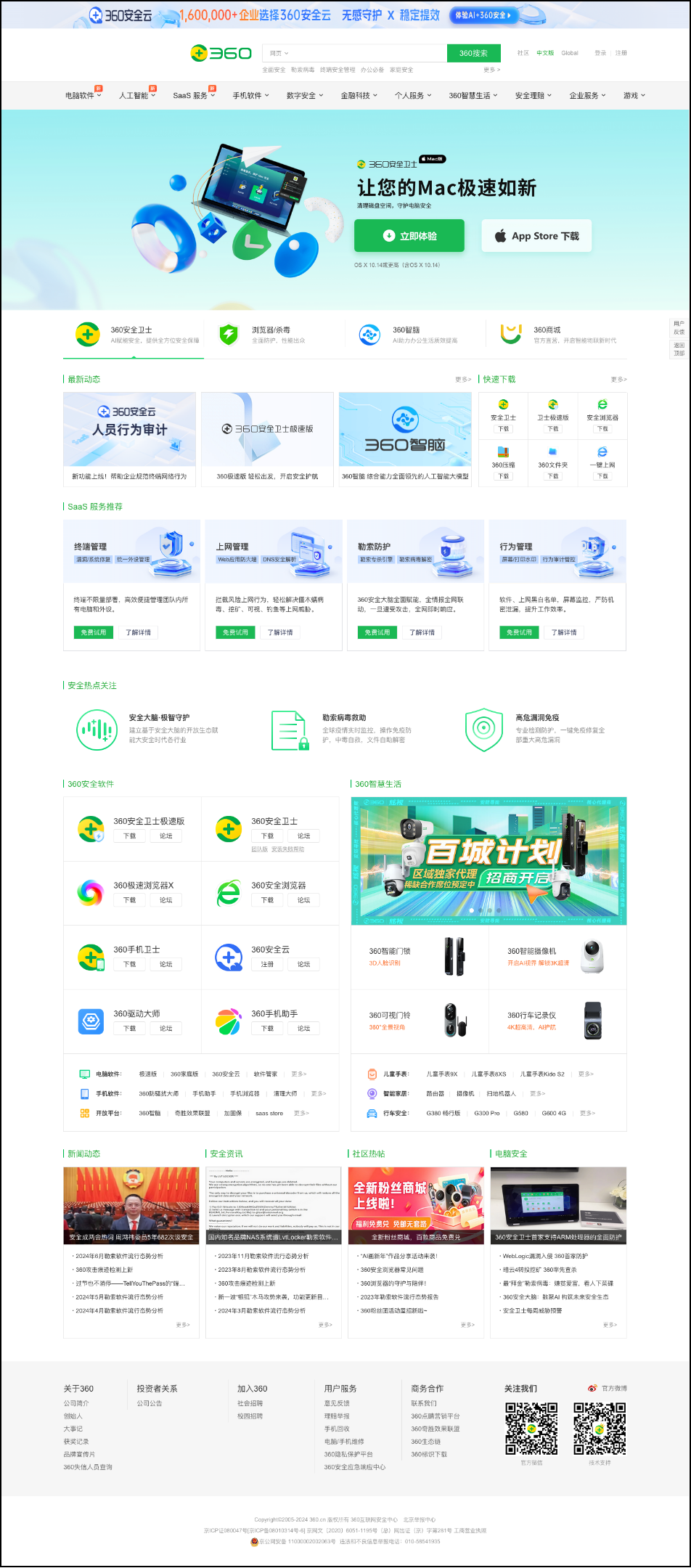}
        \caption{Reference}
        \label{fig:case_study_1_ref}
    \end{subfigure}
    \begin{subfigure}{0.25\textwidth}
        \centering
        \includegraphics[width=\linewidth]{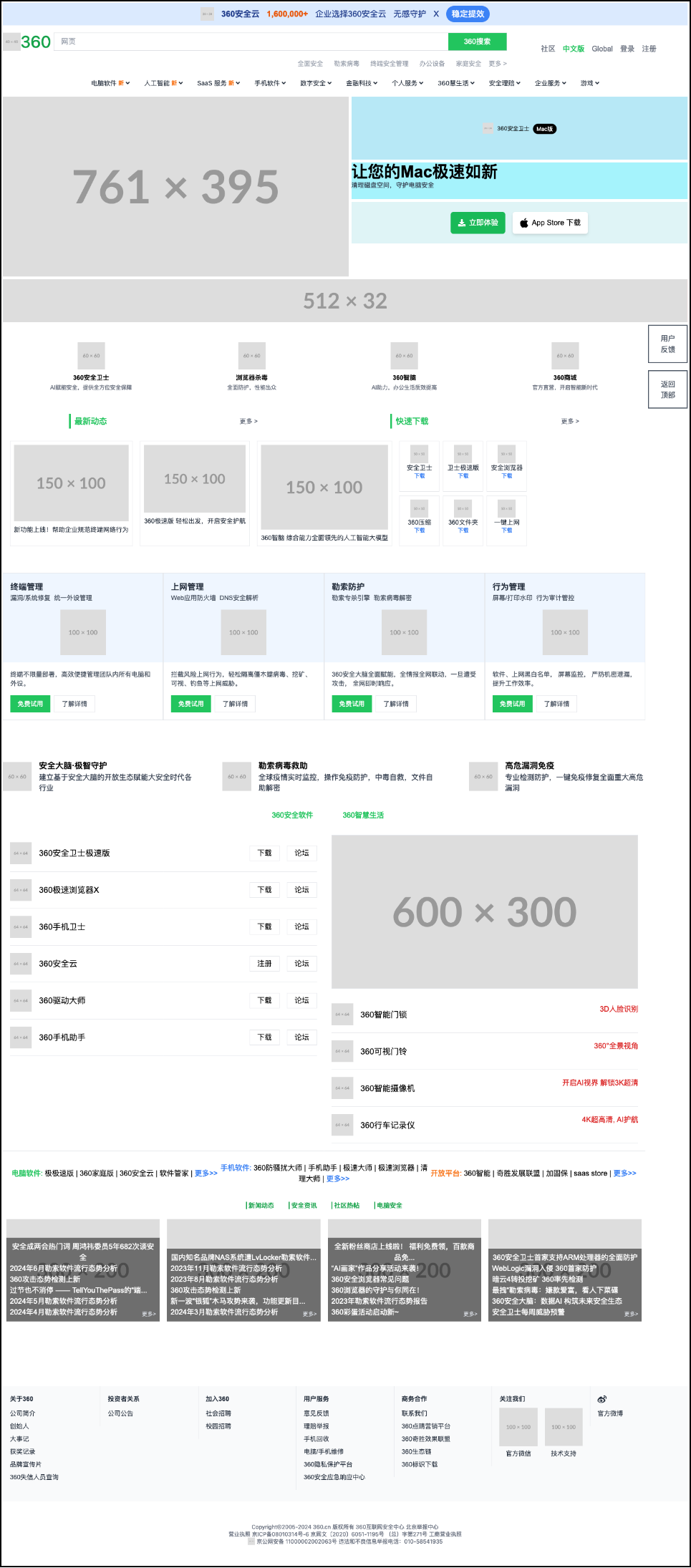}
        \caption{LayoutCoder}
        \label{fig:case_study_1_layoutcoder}
    \end{subfigure}
    \begin{subfigure}{0.25\textwidth}
        \centering
        \includegraphics[width=\linewidth]{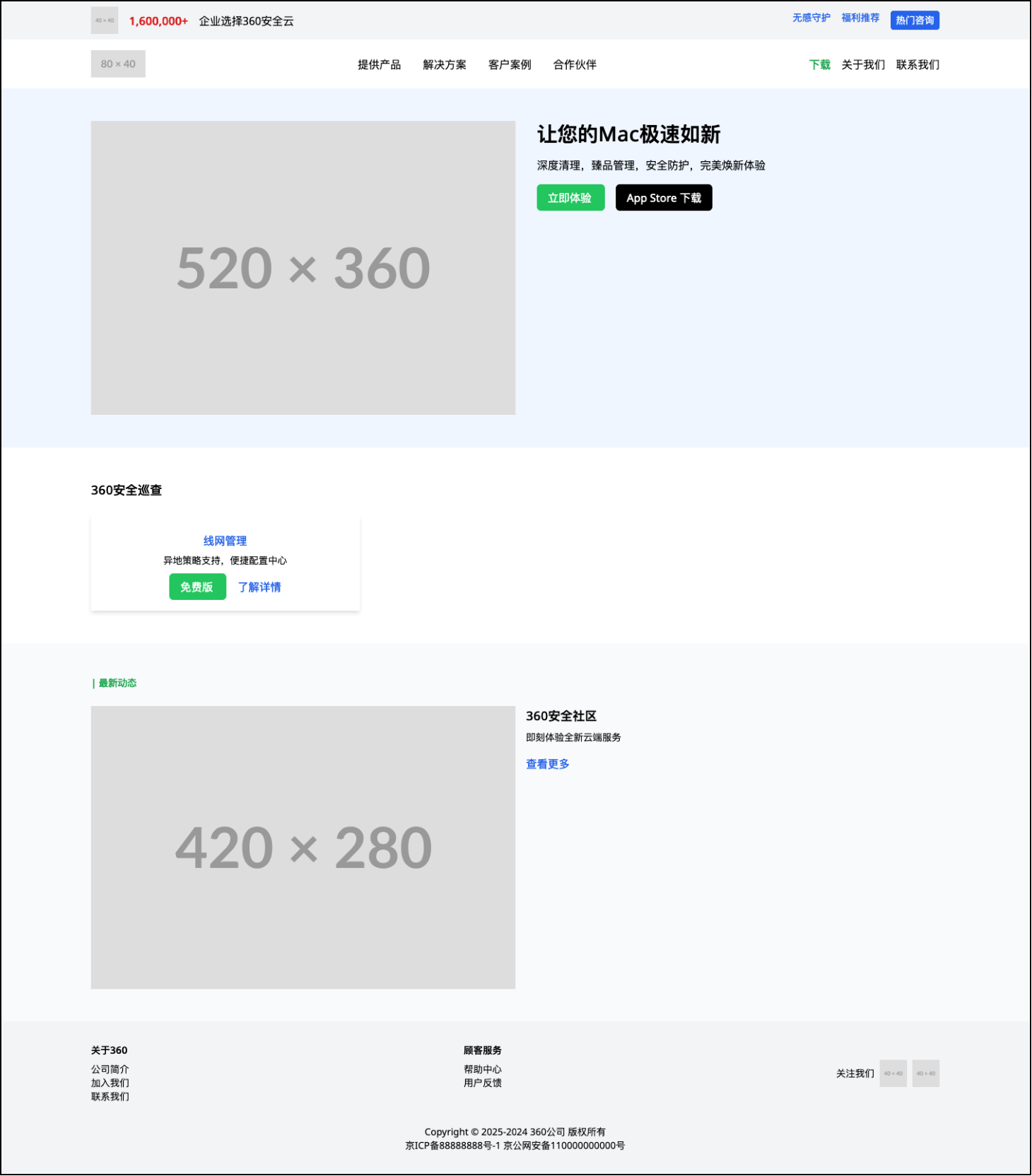}
        \caption{GPT-4o}
        \label{fig:case_study_1_gpt_4o}
    \end{subfigure}
    \vspace{-1em}
    \caption{An example illustrating the difference between the web pages generated by \tool and GPT-4o.}
    \label{fig:discussion_case_study}
    \vspace{-1em}
\end{figure}

\textbf{Observation 1: \tool better accurately captures the layout information from webpage images.}  As shown in \autoref{fig:discussion_case_study}, the first image is sampled from real web pages, while the second and third image display the rendering results of the code generated by \tool and GPT-4o, respectively. By comparing the layout and components arrangement of three images, we observe that \tool produces better results compared to GPT-4o. This observation suggests that \tool can capture layout information of the webpage image than GPT-4o.

\begin{figure}[h]
    
    \centering
    \begin{subfigure}{0.3\textwidth}
        \centering
        \includegraphics[width=\linewidth]{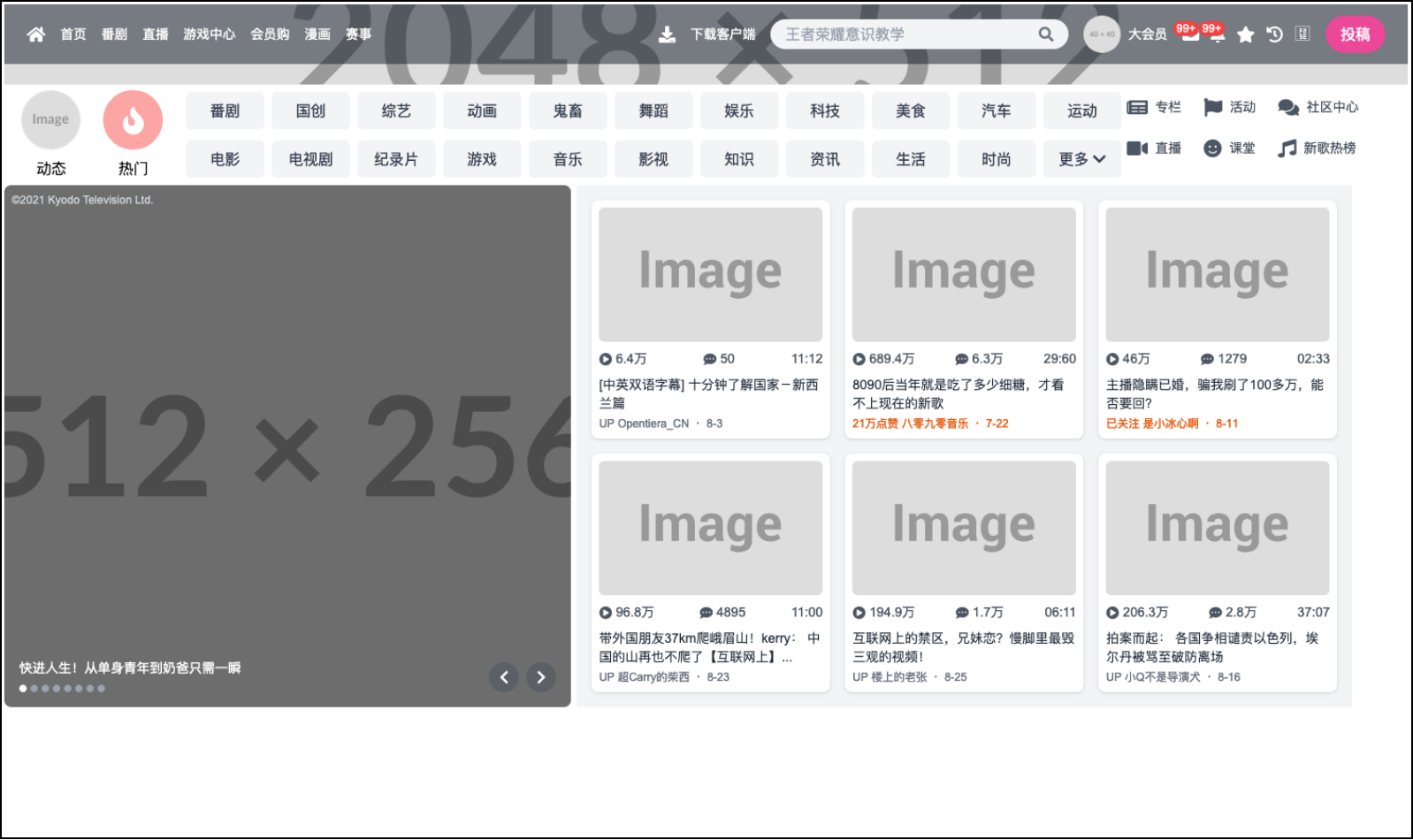}
        \caption{LayoutCoder}
        \label{fig:case_study_layoutcoder}
    \end{subfigure}
    \hfill
    \begin{subfigure}{0.3\textwidth}
        \centering
        \includegraphics[width=\linewidth]{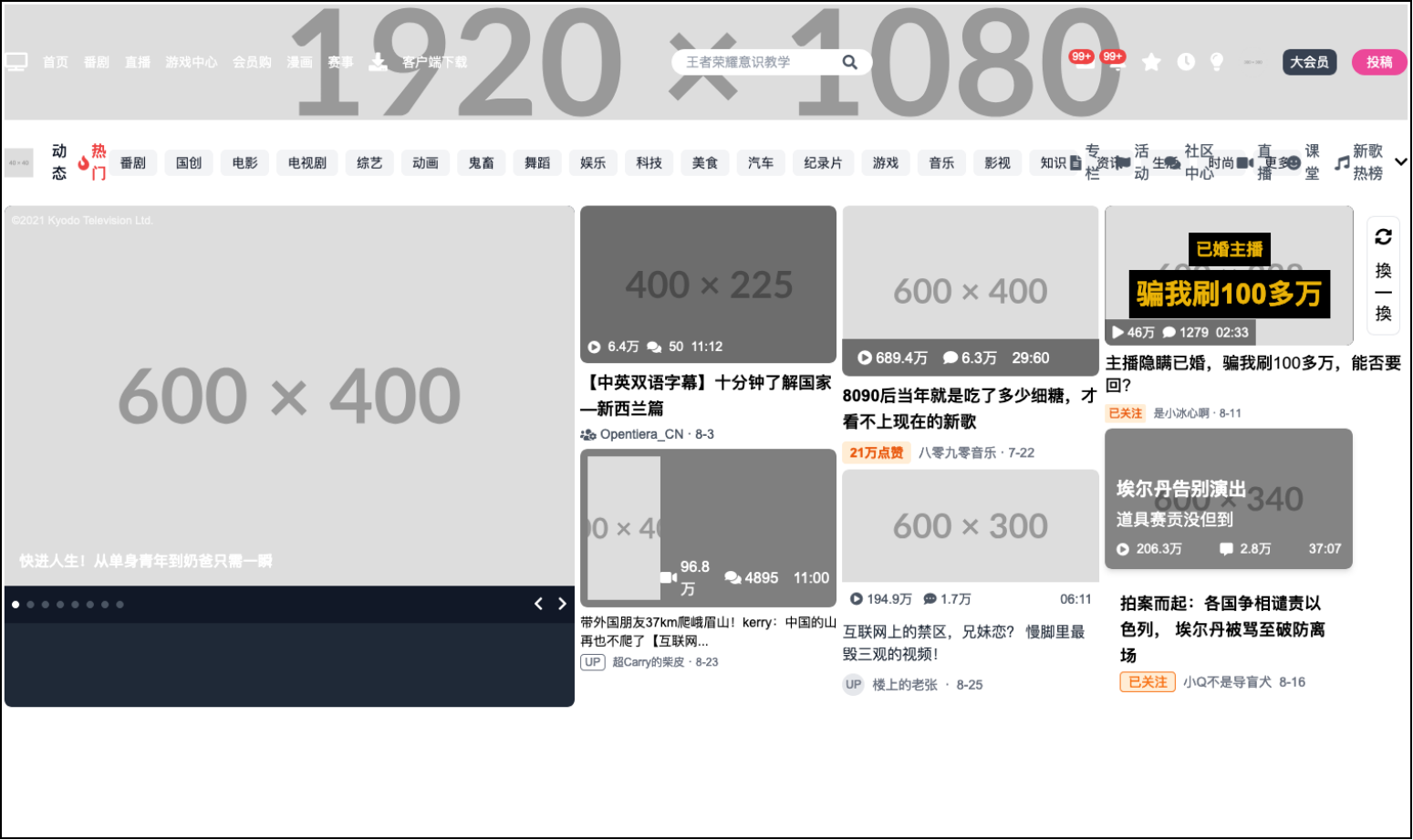}
        \caption{w/o UI Grouping}
        \label{fig:case_study_layoutcoder_wo_grouping}
    \end{subfigure}
    \hfill
    \begin{subfigure}{0.3\textwidth}
        \centering
        \includegraphics[width=\linewidth]{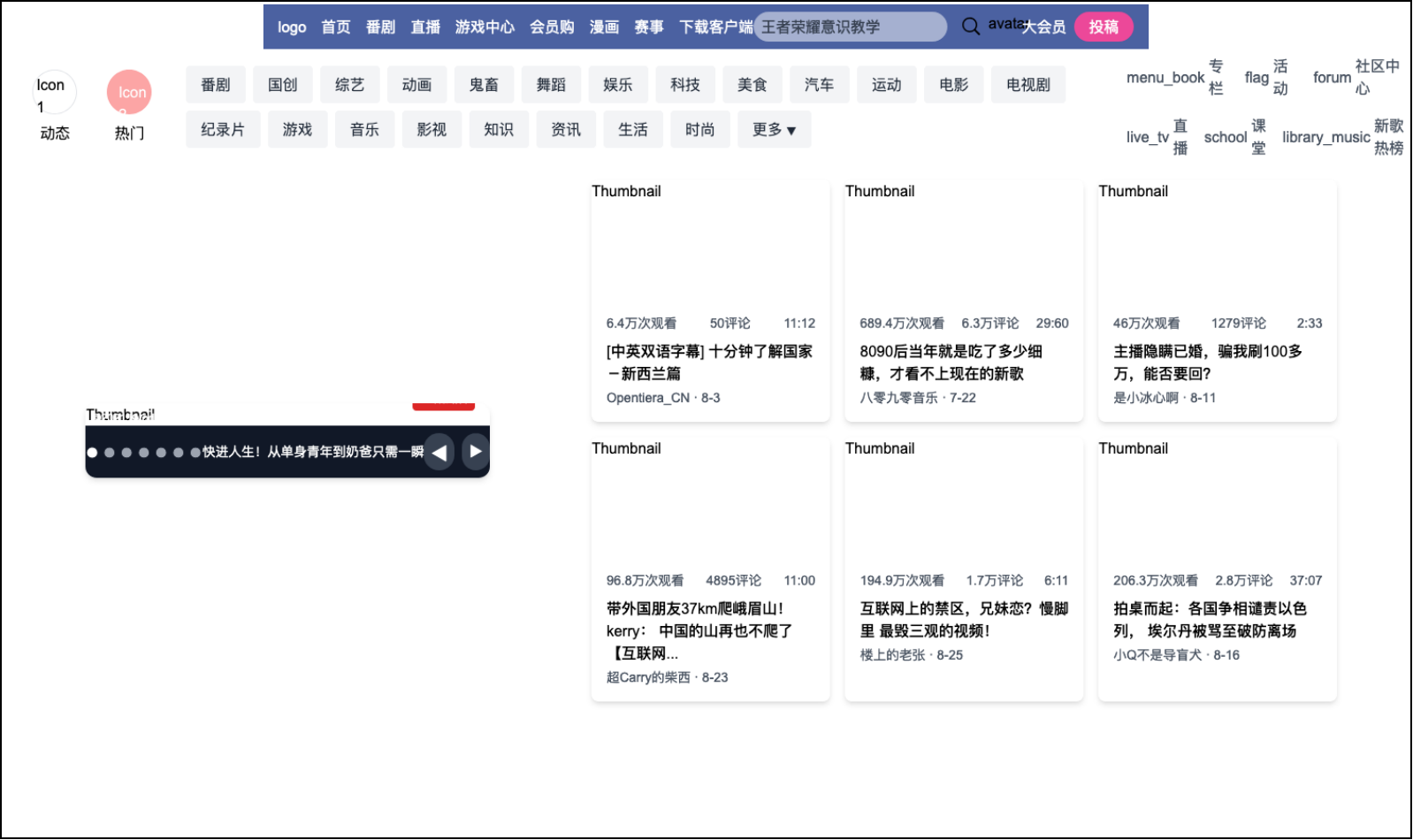}
        \caption{w/o Prompt}
        \label{fig:case_study_layoutcoder_wo_prompt}
    \end{subfigure}
    \vspace{-1em}
    \caption{Examples illustrating the differences between the web pages generated by \tool and its variants.}
    \label{fig:discussion_case_study2}
    \vspace{-1em}
\end{figure}

\textbf{Observation 2: \tool's UI grouping enhances the visual consistency of repeated components.} As shown in \autoref{fig:case_study_layoutcoder_wo_grouping}, when \tool excludes UI grouping, similar structural regions are not merged in a UI group. It leads to that six video components with identical structures exhibit visually inconsistent styles. The absence of UI grouping results in an over-segmentation of the webpage image, mistakenly treating the six video components as six independent structures. This indicates that UI grouping enhances the visual consistency of repeated components.

\textbf{Observation 3: \tool's prompt improves the structural compatibility between inner divs and outer divs.} As shown in \autoref{fig:case_study_layoutcoder_wo_prompt}, when the prompt is simplified, layout and style of inner divs conflicts with those of outer divs. This lack of the complete prompt disrupts code generation and visual rendering uniformity. This highlights the \tool's complete prompt improves the structural compatibility between inner divs and outer divs.

\subsection{Key Differences between LayoutCoder and Other Existing Methods}

\add{As shown in \autoref{tab:method_cmp}, the three methods, LayoutCoder, DCGen\cite{DCGen}, and Screen Parsing\cite{ScreenParsing}, are analyzed in terms of three key components: layout extraction, partial code generation, and code fusion.}

\add{\textbf{Key difference 1:} For Layout Extraction, Screen Parsing relies on deep learning, requiring large, high-quality training data and being UI-specific. In contrast, both DCGen and LayoutCoder are data-independent. However, DCGen struggles with complex UIs due to its reliance on line segmentation thresholds, whereas LayoutCoder uses a projection-based method, effectively handling deep UI structures.}

\add{\textbf{Key difference 2:} For Partial Code Generation, Screen Parsing does not generate code but only extracts mobile UI layouts, while both DCGen and LayoutCoder use MLLMs for code generation.}

\add{\textbf{Key difference 3:} For Code Fusion, Screen Parsing does not generate code. DCGen uses MLLMs but is limited by token-length constraints for complex webpages, whereas LayoutCoder bypasses MLLM limitations by using decoupling relationships between div elements, enabling more effective code fusion.}

\add{The primary contribution of LayoutCoder is its ability to extract layouts without relying on training data, making it highly adaptable to any webpage across both mobile and desktop platforms. Furthermore, LayoutCoder employs a non-MLLM-based code fusion approach, enabling the generation of webpages of arbitrary length, thus overcoming the token length limitations imposed by MLLMs.}

\begin{table}[t]
\caption{The key differences between LayoutCoder, DCGen and Screen Parsing.}
\label{tab:method_cmp}
\vspace{-1em}
\resizebox{\textwidth}{!}{%
\begin{tabular}{lllll}
\hline
\multirow{2}{*}{}               & \multirow{2}{*}{\textbf{Layout Extraction}}      & \multirow{2}{*}{\textbf{Partial Code Generation}}                            & \multirow{2}{*}{\textbf{Code Fusion}} & \multirow{2}{*}{\textbf{Scenarios}} \\
                                &                                                 &                                                                     &                              &                                        \\ \hline
\multirow{2}{*}{\textbf{Screen Parsing}} & \multirow{2}{*}{Deep Learning (data-dependent)} & \multirow{2}{*}{/} & \multirow{2}{*}{/}           & \multirow{2}{*}{Mobile}           \\
                                &                                                 &                                                                     &                              &                                        \\
\textbf{DCGen}                           & Separation Line Detection Algorithm           & MLLM for code generation                                                 & MLLM-based Code Fusion          & Web                \\
\textbf{LayoutCoder}                     & Projection Division Algorithm      & MLLM for code generation                                                 & Code Fusion Algorithm        & Web               \\ \hline
\end{tabular}%
}
\vspace{-1em}
\end{table}

\subsection{Comparison between LayoutCoder and DCGen}

\add{We conducted experiments on the Design2Code, Snap2Code(Seen), and Snap2Code(Unseen) datasets, recording time and token consumption during the generation process. In small-scale trials with DCGen, we observed a higher failure rate and significantly increased time and token consumption on datasets with a high Aspect Ratio. These issues stem from DCGen’s reliance on MLLM-based Code Fusion, which is limited by the MLLM's context window length.}

\add{Further analysis was conducted using a subset of 50 samples from the Snap2Code(Seen) dataset, which has a higher Aspect Ratio. For datasets with a lower Aspect Ratio, such as Design2Code and Snap2Code(Unseen), full-scale experiments were performed. As shown in \autoref{tab:dcgen_cmp}, DCGen’s time consumption on different datasets was 2-3 times higher than LayoutCoder’s, with increased token consumption and economic costs. DCGen also exhibited a 60\% success rate on Snap2Code(Seen), while LayoutCoder achieved 100\%.}

\add{To ensure fairness, we excluded failed DCGen cases from the sample. \autoref{tab:dcgen_perf_cmp} compares the performance of DCGen and LayoutCoder on two datasets. The results demonstrate that LayoutCoder outperforms DCGen in terms of time efficiency, token consumption, economic cost, code generation success rate, and overall performance across both datasets.}

\begin{table}[t]
\caption{Comparison of time efficiency, token usage, and generation success rate across two methods.}
\label{tab:dcgen_cmp}

\resizebox{\textwidth}{!}{%
\begin{tabular}{ll|rrrrrr}
\hline
\textbf{Method}                       & \textbf{Dataset}                & \multicolumn{1}{c}{\textbf{Time}} & \multicolumn{1}{c}{\textbf{Prompt Tokens}} & \multicolumn{1}{c}{\textbf{Completion Tokens}} & \multicolumn{1}{c}{\textbf{Total Tokens}} & \multicolumn{1}{c}{\textbf{Cost(Dollars)}} & \multicolumn{1}{c}{\textbf{Generation Success Rate}} \\ \hline
\multirow{3}{*}{\textbf{DCGen}}       & \textbf{Design2Code}       & 120.12                            & 9752                                       & 3870                                           & 13622                                     & 6.31                                     & 89.20\%                                               \\
                                      & \textbf{Snap2Code(Seen)}    & 1214.72                           & 31457                                      & 8989                                           & 40446                                     & 16.85                                   & 60.00\%                                                \\
                                      & \textbf{Snap2Code(Unseen)} & 398.59                            & 12314                                      & 4139                                           & 16453                                     & 7.22                                    & 91.00\%                                                \\ \hline
\multirow{4}{*}{\textbf{LayoutCoder}} & \textbf{Design2Code}       & 58.31                             & 6989                                       & 1156                                           & 8145                                      & 2.90                                  & 100.00\%                                                \\
                                    
                                      & \textbf{Snap2Code(Seen)}    & 327.50                            & 20153                                      & 5167                                           & 25320                                     & 10.21                                    & 100.00\%                                                 \\
                                      & \textbf{Snap2Code(Unseen)} & 155.41                            & 8522                                       & 1516                                           & 10037                                     & 3.65                                    & 100.00\%                                                \\ \hline
\end{tabular}%
}
\vspace{-2em}
\end{table}

\begin{table}[t]
\vspace{-1em}
\caption{Performance comparison between DCGen and LayoutCoder (excluding failures). (\%)}
\label{tab:dcgen_perf_cmp}
\vspace{-1em}
\resizebox{0.7\textwidth}{!}{%
\begin{tabular}{crrrr}
\hline
\multirow{2}{*}{Dataset} & \multicolumn{2}{c}{CLIP}                                    & \multicolumn{2}{c}{BLEU}                                    \\ \cline{2-5} 
                         & \multicolumn{1}{c}{DCGen} & \multicolumn{1}{c}{LayoutCoder} & \multicolumn{1}{c}{DCGen} & \multicolumn{1}{c}{LayoutCoder} \\ \hline
Design2Code(223)         & 79.28                     & 81.11                           & 1.94                      & 5.08                            \\
Snap2Code(Seen)(30)      & 76.76                     & 80.03                           & 3.98                      & 21.05                           \\
Snap2Code(Unseen)(91)    & 74.34                     & 75.56                           & 1.51                      & 5.70                            \\ \hline
\end{tabular}%
}
\vspace{-1em}
\end{table}

\subsection{Threats to Validity}
We identified the following three situations that pose potential threats to the effectiveness of LayoutCoder.

\subsubsection{The Selection of Baselines.}
In this paper, we present a comparison of \tool, against three baselines These baselines are all representative and demonstrate superior performance in the existing dataset.  
Due to our limited computing resources at the time of the paper writing, we do not conduct experiments with a number of other MLLMs such as LLaVA1.5. In the future, we will conduct more experiments on wider types of baselines.

\subsubsection{Generalizability on Webpages.}
The previous works use Design2Code datasets to generate  UI code from real-world webpage images. Following the previous methods \cite{REMAUI, DCGen, DeclarUI}, we limit the size of each dataset to 250 samples to control the experimental budget.  To ensure a robust evaluation, we further collet Snap2Code, retains the original images within the web pages.
It presents a more realistic and challenging scenario for the UI2Code task, aiming to improve the generalizability of \tool across various webpages.

\subsubsection{Metric Bias.}
\add{In this paper, we employ BLEU and CLIP scores to assess code similarity and visual consistency in the UI2Code task. While widely used in natural language processing and computer vision, these metrics may not fully capture the nuances of UI2Code. Due to the lack of a standardized evaluation metric for this task, our approach relies on these measures and human assessment, which introduces potential metric bias. In future work, we aim to propose a new evaluation metric specifically for UI2Code, offering a more robust and reliable framework.}

\section{Conclusion}
In this paper, we introduced \tool, a novel MLLM-based framework generating UI code from real-world webpage image, which \yun{includes} three key modules: (1) Element Relation \yun{Construction,} (2) UI Layout Parsing, and (3) Layout-Guided Code Fusion. These modules can improve the ability to understand complex layouts and generate the accurate code with layout preserved. \add{Experimental results demonstrate the superiority of \tool over the best baseline, Claude 3.5 Sonnet in both visual and code similarity on real-world web pages and robust generalization on highly complex web pages.}

\section*{Data Availability}

The artifacts are archived at this link~\cite{layoutcoder_github}.

\begin{acks}
This research is supported by National Key R\&D Program of China (No. 2022YFB3103900), National Natural Science Foundation of China under project (No. 62472126), Natural Science Foundation of Guangdong Province (Project No. 2023A1515011959), Shenzhen-Hong Kong Jointly Funded Project (Category A, No. SGDX20230116091246007), Shenzhen Basic Research (General Project No. JCYJ20220531095214031), Shenzhen International Science and Technology Cooperation Project (No. GJHZ20220913143008015), the Major Key Project of PCL (Grant No. PCL2024A05), and CCF-Huawei Populus Grove Fund.
\end{acks}

\bibliographystyle{ACM-Reference-Format}

\normalem
\bibliography{ref}

\appendix

\end{document}